\documentclass[aps,a4paper, preprint, superscriptaddress,preprintnumbers,floatfix,nofootinbib,amsmath,amssymb]{revtex4}
\usepackage{url}
\usepackage{hyperref}
\usepackage{cleveref}

\usepackage{amstext,amssymb}
\usepackage[section]{placeins}
\usepackage{amsmath}
\usepackage{graphicx}
\usepackage{xspace}
\usepackage{color}
\usepackage{xcolor}
\usepackage{float}
\usepackage{comment}
\usepackage{amstext,amssymb}
\usepackage{amsmath}
\usepackage{graphicx}
\usepackage[section]{placeins}
\usepackage{xspace}
\usepackage{color}
\usepackage{units}
\usepackage{array}
\usepackage[section]{placeins}
\usepackage{amsmath,bm}
\usepackage{amssymb}
\usepackage{gensymb}
\usepackage{soul}
\usepackage{textcomp}
\usepackage{slashed} 
\usepackage{multirow}
\usepackage{hyperref}
\usepackage{mathrsfs}
\usepackage{enumitem}
\usepackage{graphicx}
\usepackage{color}
\usepackage{comment}
\usepackage{epstopdf}
\usepackage{float}
\usepackage{units}
\usepackage{slashed}
\usepackage{amsmath,bm}
\usepackage{slashed} 
\usepackage{multirow}
\newcommand{\be}{\begin{equation}}
\newcommand{\ee}{\end{equation}}
\newcommand{\bea}{\begin{eqnarray}}
\newcommand{\eea}{\end{eqnarray}}

\def\s1{\hat s}

\usepackage{slashed}

\newcommand{\nua}[1]{\ensuremath{\rlap{\kern-2.5pt\ensuremath{\overset{\scriptscriptstyle(-)}{\phantom{\nu}}}}{\ensuremath{{\nu}_{#1}}}}\xspace}

\large
\begin{document}
\title{Discernible NSI Effects in Long-Baseline Neutrino Experiments}


\author{Barnali Brahma}
\email{ph19resch11001@iith.ac.in}

\author{Anjan Giri}
\email{giria@iith.ac.in}

\affiliation{Department of Physics, IIT Hyderabad,
              Kandi - 502285, India }      




\begin{abstract}
{Neutrino oscillation in the matter could get affected by the sub-dominant, yet unknown, non-standard interactions. The upcoming  long-baseline (LBL)  neutrino experiments will be sensitive to these effects and can provide information on the unknown oscillation parameter values. In this article, we study the parameter degeneracies that can occur in DUNE, T2HK experiments, and a combination of both due to nonstandard interactions (NSI), arising simultaneously, from two different off-diagonal sectors, i.e., $e-\mu$ and $e-\tau$. We derive constraints on both the NSI sectors using the combined datasets of NO$\nu$A and T2K. Our analysis reveals a significant impact that dual NSIs may have on the sensitivity of atmospheric mixing angle $\theta_{23}$ in the normal ordering (NO) case.  Furthermore, when non-standard interaction from the $e-\mu$ and $e-\tau$ sectors are included, we see significant changes in the probabilities for DUNE, T2HK, and as well as a combined analysis
involving both. Moreover, the CP sensitivity gets affected significantly due to the presence of dual NSIs, and, in addition, the CP asymmetry also exhibits an appreciable difference.}
\end{abstract}
\maketitle


\section{Introduction}
{The success of  neutrino oscillation experiments in the last decades has been a major achievement in modern physics. The results of several neutrino investigations employing solar, atmospheric, reactor, and accelerator neutrinos can be explained using the three neutrino mixing model, in which the three known neutrino flavor eigenstates ($\nu_e, \nu_{\mu}, \nu_{\tau}$) are quantum superpositions of three mass eigenstates ($\nu_1, \nu_2, \nu_3$), making them one of the purest probes of quantum mechanics known. Six oscillation parameters dictate the neutrino oscillation probabilities in the SM with three neutrinos: two mass-squared differences ($\Delta m^{2}_{21}, \Delta m^{2}_{31}$), three mixing angles ($\theta_{12},\theta_{13},\theta_{23}$), and one Dirac CP phase $\delta_{CP}$. In front of the current and upcoming neutrino experiments,  there are three important challenges: $\theta_{23}$ octant  degeneracy, the value of  CP-violating phase $\delta$, and sgn ($\Delta m^{2}_{31}$).\par
In recent times, neutrino oscillation programs  have been able to  measure the known parameters with ever-increasing accuracy ~\cite{Super-Kamiokande:1998kpq, SNO:2002tuh, KamLAND:2004mhv, MINOS:2011neo}. The effort of many dedicated neutrino oscillation experiments over the last two decades provides us with a splendid understanding of the main features of these tiny and elusive particles. The upcoming long-baseline neutrino experiments  will also be able to determine the subdominant effects present in the oscillation data, which are sensitive to the yet unknown oscillation parameter values.  All of the physics discussed here is based on the framework of neutrino mass and mixing concept ~\cite{Minakata:2001qm,Fogli:1996pv}. The neutrinos interact with matter through the SM weak interaction and gravity. The interaction due to gravity  is so weak that it is irrelevant for inclusion or discussion.\par
On entering the Earth's atmosphere and traveling through the Earth's crust, the neutrino gets influenced by a matter potential, known as Wolfenstein's matter effect. To probe new physics, Wolfenstein introduced non-standard interaction besides the neutrino mass matrix \cite{Wolfenstein:1977u}. There have been  extensive studies of neutrino phenomenology in the literature ~\cite{Davis:1968cp, Minkowski:1977sc, Mohapatra:1979ia, Guzzo:1991hi, Gonzalez-Garcia:1998ryc, Friedland:2005vy, Miranda:2004nb, Machado:2011ug, Esteban:2020cvm}. At low energy, NSI describes new physics models where the neutrino interaction with ordinary matter is parameterized in terms of effective couplings, $\epsilon_{\alpha \beta}$, where $\alpha$ and $\beta$ are the generation indices. ~\cite{Barger:1991ae, Grossman:1995wx, Gonzalez-Garcia:2011vlg, Huber:2001zw, Fornengo:2001pm, Blennow:2007pu, Palazzo:2009rb, Coloma:2011rq, Esmaili:2013fva, MINOS:2013hmj, Choubey:2015xha, Miranda:2015dra, Bakhti:2016prn, Capozzi:2013csa, Forero:2014bxa,Babu:2019vff, Majhi:2022wyp}. NSI arises naturally in neutrino mass models in trying to explain the large    \cite{ Proceedings:2019qno, Ohlsson:2012kf} as well as   small \cite{Farzan:2015hkd, Farzan:2017xzy, Farzan:2018gtr} neutrino mixing angles. Hence, it is quite crucial to understand the effect of NSI on standard neutrino oscillation in the matter. In general, NSI affects neutrino propagation in a matter not only through neutral-current interactions but also through charged-current interactions, which affects neutrino generation and detection. Model-independent bounds on the production and detection of NSI are typically an order of magnitude higher than those of the matter NSI \cite{Biggio:2009nt}, and therefore, 
we ignore the production and detection of NSI in this study and concentrate only on matter NSI.\par
The standard model (SM) CP phase promises to help us comprehend the universe's baryon asymmetry and is the most sought-after observable in the present and future neutrino investigations. The latest data from the two long-baseline accelerator studies, NO$\nu$A \cite{NOvA:2021nfi} and T2K \cite{T2K:2021xwb} show some tension in the usual three-flavor scenario. NO$\nu$A prefers the CP phase to be close to $\delta_{CP} \approx$ 0.8$\pi$ whereas T2K hints at a value of $\delta_{CP}$ around  $1.5\pi$ in the case of normal ordering. There appears to be no disagreement in the case of inverted ordering. Once the NSI from the $e-\mu$ sector is taken into account the tension concerning the $\delta_{CP}$ parameter for NO$\nu$A and T2K becomes tranquil, but one can see a difference for $\theta_{23}$ ~\cite{Chatterjee:2020kkm, Denton:2020uda}. NO$\nu$A prefers a lower octant, whereas T2K prefers a higher octant. We have explored the degeneracy issue for the standard model parameter $\theta_{23}$ in the presence of NSI arising individually from both $e-\mu$ and $e-\tau$ sectors for DUNE and T2HK in the Ref ~\cite{Brahma:2022xld}.

In this paper, we review the dual NSI effect arising from $e-\mu$ and $e-\tau$ sectors at once and show how the NSI parameters can affect the SM oscillation parameters in the upcoming long-baseline neutrino experiments. We retrieved NO$\nu$A ~\cite{nova} and T2K \cite{t2k} datasets from the most recent data release in order to determine the constraints on NSI contributions. The same coefficients are then used to examine if we can obtain any perceptible result in future long-baseline (LBL) neutrino experiments, such as the DUNE and T2HK. It's worth noting that, the various baselines and energy of T2HK and DUNE are responsible for differences in terms of their sensitivity to matter effects. This raises the intriguing prospect that new physics in the form of non-standard neutrino interactions could be at work. Here, we show how the sensitivity change for SM, with respect to the dual NSIs  effects, in T2HK, DUNE, and a combination of both of these experiments.}

{\section{Formalism:}}
{The NSI is defined by dimension six four-fermion ($ff$) operators of the form ~\cite{Wolfenstein:1977u}:
\begin{equation}\label{1}
    {\mathcal{L}}_{NSI} = 2\sqrt{2}G_{F} \epsilon_{\alpha \beta}^{fC} [  \overline{\nu_{\alpha}} \gamma^{\rho} P_{L}  \nu_{\beta}][\overline{f} \gamma_{\rho} P_{C} f] + h.c.
\end{equation}
where $G_{F}$ is Fermi coupling constant, $\epsilon_{\alpha \beta}^{fC}$ are dimensionless parameters that measure the new interaction's strength in relation to the SM, $\alpha, \beta= e, \mu, \tau$ indicate the neutrino flavor, superscript $C = L, R$ refers to the chirality of $ff$ current, $f = u, d, e $ denotes the matter fermions. 
The neutrino propagation Hamiltonian in the presence of matter, NSI, can be expressed as

\begin{align*}
H_{Eff} = \frac{1}{2E}
\Bigg[ U_{PMNS}
\begin{bmatrix}
 0 & 0 & 0 \\
0 & \Delta{m^{2}_{21}} & 0\\
0 &0 & \Delta{m^{2}_{31}} \\
\end{bmatrix}
 U^{\dagger}_{PMNS}
 + V\Bigg]
\end{align*}
where unitary Potecorvo-Maki-Nakagawa-Sakata mixing matrix is denoted by $U_{PMNS}$, neutrino energy as E, the different mass eigenstates as $m_{1}$, $m_{2}$ and $m_{3}$ and $\Delta m_{21}^{2}\equiv\ m_{2}^{2}-m_{1}^{2}$, $\Delta m_{31}^{2}\equiv m_{3}^{2}-m_{1}^{2}$. $V$ is written as:

\begin{equation*}
V = 2\sqrt{2}G_{F}N_{e}E
\begin{bmatrix}
  1+\epsilon_{ee}& \epsilon_{e \mu}e^{i \phi_{e \mu}} & \epsilon_{e \tau}e^{i \phi_{e \tau}} \\
  \epsilon_{ \mu e}e^{-i \phi_{e \mu}}  & \epsilon_{\mu \mu} & \epsilon_{\mu \tau}e^{i \phi_{\mu \tau}} \\
  \epsilon_{\tau e} e^{-i \phi_{e \tau}} & \epsilon_{\tau \mu}e^{-i \phi_{\mu \tau}} & \epsilon_{\tau \tau}\nonumber\\
\end{bmatrix}
\end{equation*}

$N_{e}$ is the number density of electrons and for neutrino propagation in the Earth, $\epsilon_{\alpha\beta}e^{i\phi_{\alpha \beta}} \equiv \sum_{f,C}\epsilon_{\alpha\beta}^{fC} \frac{N_{f}}{N_{e}} \equiv \sum_{f=e,u,d}(\epsilon_{\alpha\beta}^{fL}+\epsilon_{\alpha\beta}^{fR}) \frac{N_{f}}{N_{e}}$,

$N_f$ being the number density of $f$ fermion. The $\epsilon_{\alpha\beta}$ are real, and $\phi_{\alpha\beta}$ = 0 for $\alpha = \beta$.
We concentrate on flavour non-diagonal NSI ($\epsilon_{\alpha \beta}$'s with $\alpha \neq \beta$). Here, we focus on the dual NSI parameter $\epsilon_{e \mu}$ and $\epsilon_{e \tau}$ (simultaneously) to examine the conversion probability of $\nu_{\mu} \rightarrow \nu_{e}$ for the LBL studies which can be stated as the sum of four (plus higher order; cubic and beyond) terms in the presence of dual NSI : 
\begin{equation}
P_{\mu e} = P_{SM} + P_{\epsilon_{e\mu}} + P_{\epsilon_{e\tau}} + P_{Int} + h.o.
\end{equation}
the above Eq.(2), similar to ~\cite{Liao:2016hsa, Kikuchi:2008vq, Kopp:2007ne, Meloni:2009ia}
takes the following form:\\
\begin{equation*}
P_{SM} = 4s_{13}^{2}s_{23}^{2}f^{2}+8s_{13}s_{23}s_{12}c_{12}c_{23}rfg\cos({\Delta+\delta_{CP}})+4r^{2}s_{12}^{2}c_{12}^{2}c_{23}^{2}g^{2}
\end{equation*}
\begin{eqnarray}
P_{\epsilon_{e\mu}}&=& 4\hat{A}\epsilon_{e\mu} [xf^{2}s^{2}_{23}\cos({\Psi_{e\mu}})+c^{2}_{23}g\cos({\Delta+\Psi_{e\mu}}) + yg^{2}c^{2}_{23}\cos{\Phi_{e\mu}} \nonumber\\
&+& ygfs^{2}_{23}\cos({\Delta - \Phi_{e\mu}})]
 +4\hat{A}^{2}\epsilon_{e\mu}^{2}[fgs_{23}c^{3}_{23}+f^{2}s_{23}^{4}+2fgs_{23}^{2}c^{2}_{23}\cos{\Delta}]\nonumber
 \end{eqnarray}
 \begin{eqnarray}
 P_{\epsilon_{e\tau}}&=& 4\hat{A}\epsilon_{e\tau} xf^{2}s_{23}c_{23}\cos({\Psi_{e\tau}}) - 4\hat{A}\epsilon_{e\tau} s_{23}c_{23}g\cos({\Delta+\Psi_{e\tau}}) - 4\hat{A}\epsilon_{e\tau} yg^{2}s_{23}c_{23}\cos{\Phi_{e\tau}} \nonumber\\
 &+& 4\hat{A}\epsilon_{e\tau}ygs_{23}
c_{23}f\cos({\Delta-\Phi_{e\tau}}) + 4\hat{A}^{2}fgs^{3}_{23}c_{23}\epsilon_{e\tau}^{2} + 4\hat{A}^{2}f^{2}s_{23}^{2}c^{2}_{23}\epsilon_{e\tau}^{2}\nonumber\\
&-& 8\hat{A}^{2}fgs_{23}^{2}c^{2}_{23}\cos{\Delta}\epsilon_{e\tau}^{2}\nonumber
\end{eqnarray}
\begin{eqnarray}
P_{Int}&=& 8\hat{A}^{2}g^{2}s_{23}c^{3}_{23}\epsilon_{e\mu}\epsilon_{e\tau} + 8\hat{A}^{2}f^{2}s^{3}_{23}c_{23}\epsilon_{e\mu}\epsilon_{e\tau} + 8\hat{A}^{2}fgs_{23}c_{23}\epsilon_{e\mu}\epsilon_{e\tau}[2c_{23}\cos(\phi_{e\mu}-\phi_{e\tau})\nonumber\\
    &-&\cos{(\Delta-\phi_{e\mu}+\phi_{e\tau})}] \nonumber
\end{eqnarray}
where, $\Delta=\frac{\Delta m^{2}_{31}L}{4E}$; $r=\frac{\Delta m^{2}_{21}}{\Delta m^{2}_{31}}$; $\hat{A}=\frac{2\sqrt{2}G_{F}N_{e}E}{\Delta m^{2}_{31}}$; $g \equiv \frac{\sin{\hat{A}\Delta}}{\hat{A}}$; ${f\equiv \frac{\sin{[(1-\hat{A})\Delta]}}{1-\hat{A}}}$. Furthermore, here we used: $\Psi_{e\mu}=\phi_{e\mu}+\delta_{CP}$; $\Psi_{e\tau}=\phi_{e\tau}+\delta_{CP}$.\\

For anti-neutrino probability, $P \equiv P (\overline{\nu}{_{e}} \rightarrow \overline{\nu}{_{\mu}} )$, is given by
changing the above expression for $P_{SM}$, $P_{\epsilon_{e\mu}}$, $P_{\epsilon_{e\tau}}$, $P_{Int}$  with $\hat{A} \rightarrow -\hat{A}$ (and hence $f \rightarrow \overline{f}$), $\delta \rightarrow -\delta$, and
$\phi_{\alpha\beta} \rightarrow -\phi_{\alpha\beta}$. For the inverted hierarchy (IH), $\Delta_{CP} \rightarrow -\Delta_{CP}$, $y \rightarrow -y$, $\hat{A} \rightarrow -\hat{A}$ (i.e., $f \leftrightarrow -\overline{f}$, and $g \leftrightarrow -g$).}

{\section{Analysis details and results:}}

{For our analysis purpose, we used DUNE and T2HK running for 3.5 years and 3 years in $\nu$ mode and similarly 3.5 years and 4 years in $\Bar{\nu}$ mode, respectively.
DUNE ~\cite{DUNE:2020ypp} will have a 40-kiloton liquid argon detector, which will use a 1.2 MW proton beam to generate neutrino and antineutrino beams from in-flight pion decays. The proton beam will originate 1300 kilometers upstream at Fermilab. The energy ranges for neutrinos will be between 0.5 and 20 GeV, with a flux peak of around 3.0 GeV. The T2HK experiment, on the other hand, will use a 225 kt water Cherenkov detector. It will employ an enhanced 30 GeV J-PARC beam with a power of 1.3 MW, with its detector located 295 kilometers from the source.\\
We use the software GLoBES ~\cite{GLoBES, Huber:2004ka, Huber:2007ji} and its supplementary public tool, which takes into account the non-standard interactions ~\cite{J.Kopp} in our analysis. 
The extension enables non-standard neutrino interactions and sterile neutrinos in GLoBES simulation. 
The best-fit values of the standard model parameters, as well as their associated uncertainties, are obtained from nuFIT v5.1 ~\cite{nufit} and PDG ~\cite{ParticleDataGroup:2022pth} as listed out in table 1. 
\begin{table}[h!]
\caption{\label{tab:table1}{\large{Parameter values }}}
\begin{center}
\begin{ruledtabular}
\begin{tabular}{||c|c|c||}
\hline
 \large{\textbf{SM Parameters}} &  \textbf{bfp $\pm$ 1$\sigma$} & \textbf{bfp $\pm$ 1$\sigma$} \\ 
   &   \textbf{NO }  & \textbf{IO} \\[1ex]  \hline
$\sin^{2}\theta_{12}$ & $0.304^{+0.012}_{-0.012}$ & $0.304^{+0.013}_{-0.012}$ \\ 
$\sin^{2}\theta_{23}$ & $0.450^{+0.019}_{-0.016}$ & $0.570^{+0.019}_{-0.016}$\\
$\sin^{2}\theta_{13} $ & $0.02246^{+0.00062}_{-0.00062}$ & $0.02241^{+0.00074}_{-0.00062}$\\
$\delta_{CP}/^{\circ}$ & $230^{+36}_{-25}$   & $278^{+22}_{-30}$\\
$\frac{\Delta m^{2}_{21}}{10^{-5} eV^{2}}$ & $7.42^{+0.21}_{-0.20}$ & $7.42^{+0.21}_{-0.20}$ \\
$\frac{\Delta m^{2}_{3l}}{10^{-3}eV^{2}}$& $+2.510^{+0.027}_{-0.027}$   & $-2.490^{+0.026}_{-0.028}$\\[1ex]
\hline
\end{tabular}
\end{ruledtabular}
\end{center}
\end{table}
We use GLoBES to combine the extracted datasets of T2K and NO$\nu$A. The sensitivity as well as the oscillation probabilities for the two next-generation LBL experiments, DUNE, T2HK, and a combination of both of these experiments, are discussed using the NSI coefficients thus acquired. We use pre-defined AEDL (a comprehensive abstract experiment definition language) files available for simulating experiments like T2HK  and DUNE.

In Fig. \ref{fig:image1}, the results of the analysis for the combination of T2K and NO$\nu$A are displayed. The left panel shows the allowed region in the plane spanned by the NSI parameters $\epsilon_{e\mu}$ and $\epsilon_{e\tau}$, whereas the right panel displays the allowed region for NSI phases $\phi_{e\mu}$ and  $\phi_{e\tau}$ for NO scenario (top panel) and IO scenario (bottom panel). For the left panel plots, $\theta_{13}$, $\delta_{CP}$ along with the non-standard CP-phases $\phi_{e\mu}$, and $\phi_{e\tau}$, are marginalized away whereas for the right panel plots, $\theta_{13}$, $\theta_{23}$ along with the non-standard magnitudes $\epsilon_{e\mu}$, and $\epsilon_{e\tau}$ are marginalized.

From the top and bottom panels of Figs. \ref{fig:image1}, we can visualize that both in NO as well as in IO cases
there is a preference for a non-zero value of the coupling $|\epsilon_{e\mu}$ and $\epsilon_{e\tau}$ and their corresponding phases $\phi_{e\mu}$ and $\phi_{e\tau}$, whose values are listed out in Table 2. These values are consistent with the global constraints on neutral current NSI parameters ~\cite{Coloma:2019mbs}. 

\begin{table}[h!]
\caption{\label{tab:table2}{From allowed region plots, the best-fit points are listed here. The best-fit points are picked up corresponding to the minimum $\chi^{2}$ value. These values are also included in the below table.}}
\begin{center}
\begin{ruledtabular}
\begin{tabular}{cccc}
 \large{Mass ordering} &  $|\epsilon_{e \mu}|$ & $|\epsilon_{e \tau}|$ & $\chi^{2}$  \\ [2ex]  \hline
NO  & 0.1  & 0.033 & 0.659  \\ 
IO  & 0.1 & 0.02 & 1.14\\ [1ex]
\hline
 \large{Mass ordering} & $\phi_{e\mu}/\pi$ & $\phi_{e \tau}/\pi$ & $\chi^{2}$  \\ [2ex]\hline
NO  & 1.06  & 1.87 & 0.549 \\
IO  & 1.0 & 1.73 &  0.952\\ 
\end{tabular}
\end{ruledtabular}
\end{center}
\end{table}
\begin{figure}[hbt!]
\minipage{0.44\textwidth}
  \includegraphics[width=\linewidth]{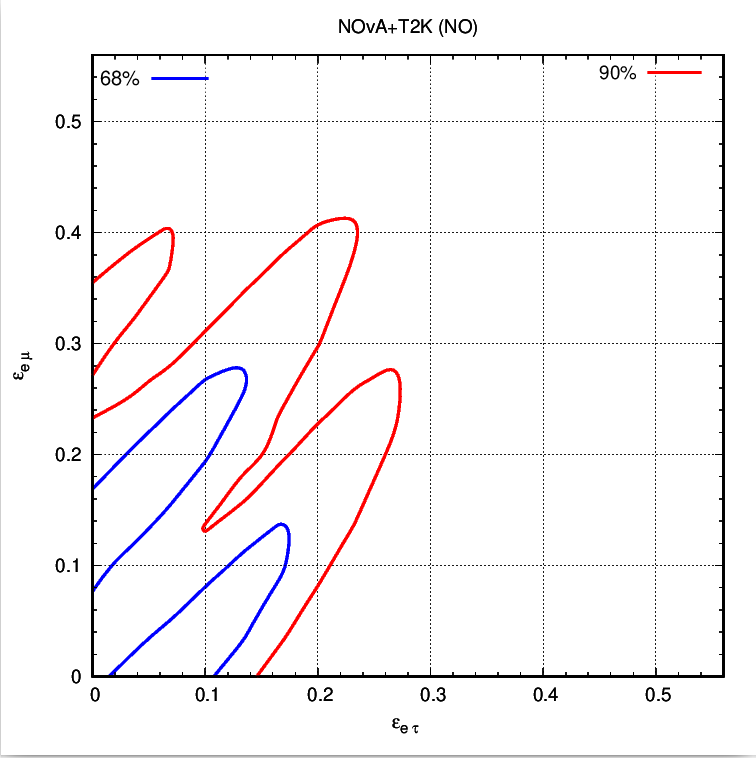}
\endminipage\hfill
\minipage{0.44\textwidth}
  \includegraphics[width=\linewidth]{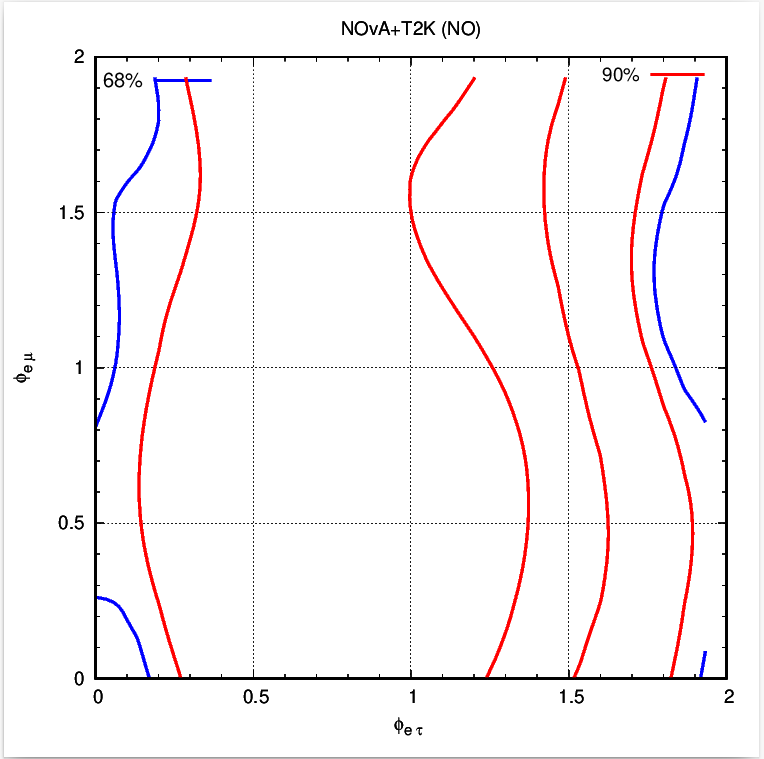}
\endminipage

\minipage{0.44\textwidth}
  \includegraphics[width=\linewidth]{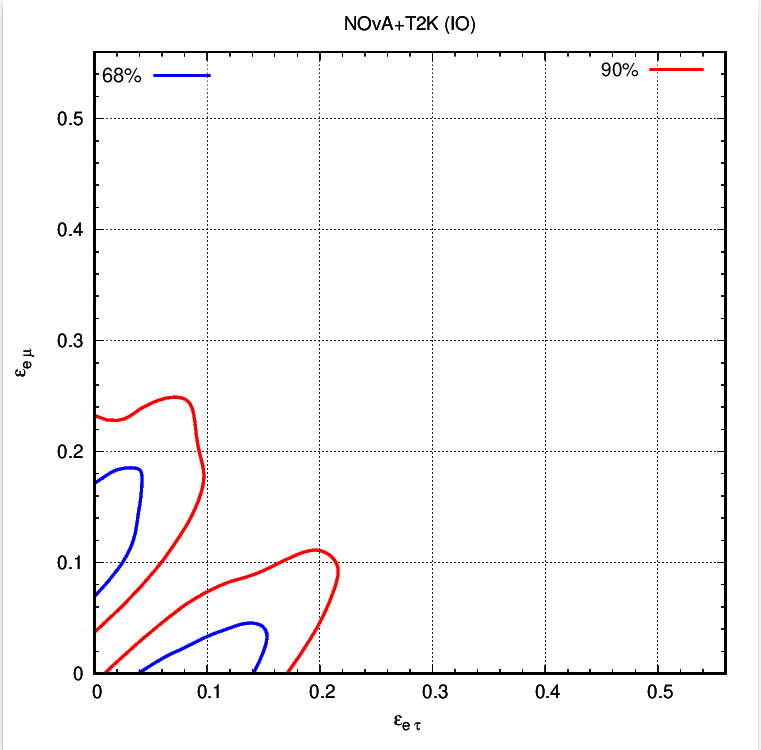}
\endminipage\hfill
\minipage{0.44\textwidth}
  \includegraphics[width=\linewidth]{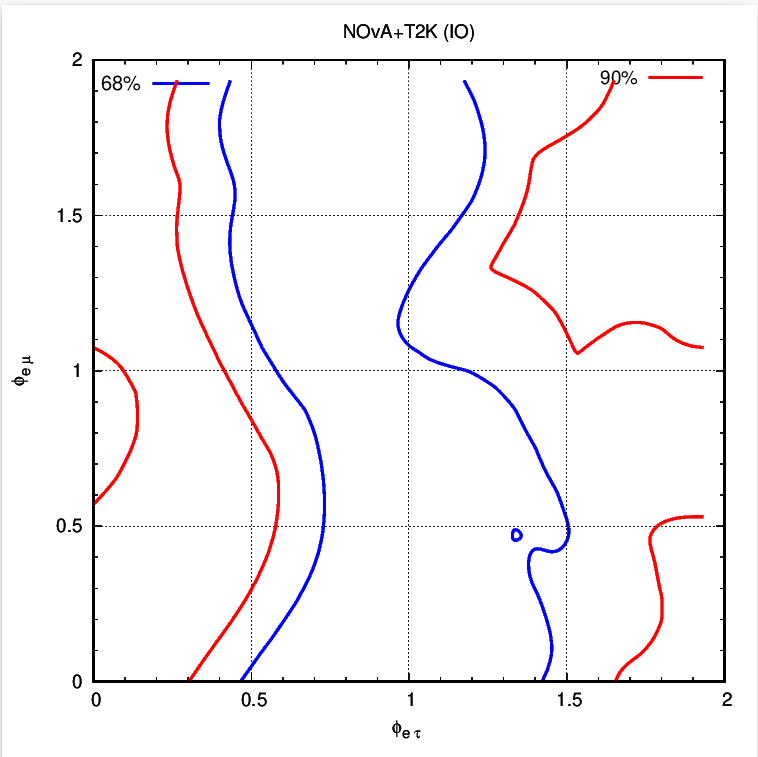}
\endminipage
\caption{Allowed regions 
in the plane spanned by NSI coupling 
for $\epsilon_{e\mu}$ and $\epsilon_{e\tau}$ (left); $\phi_{e\mu}$ and phase $\phi_{e \tau}$(right) determined by the combination of T2K and NO$\nu$A for NO (top panel) and IO(bottom panel). The contours are drawn at the 68$\%$ and 90$\%$ C.L. for 2 d.o.f}\label{fig:image1}
\end{figure}

In Fig. \ref{fig:image2}, we display the allowed regions in the plane spanned by the standard CP-phase $\delta_{CP}$ and the atmospheric mixing angle $\theta_{23}$ in the NO case for DUNE (top panel). The left panel refers to the SM case, while the right panel concerns the SM along with the dual NSI scenario, arising from the $e-\mu$ and $e-\tau$ sectors, taken simultaneously. The mixing angle $\theta_{13}$ and $\Delta m^{2}_{31}$  are marginalized away in the SM case whereas along with $\theta_{13}$ and $\Delta m^{2}_{31}$, relevant NSI couplings ($\epsilon_{e\mu}$ and $\epsilon_{e\tau}$) and non-standard CP-phases ($\phi_{e\mu}$ and $\phi_{e\tau}$) are marginalized in SM+2NSI case. We have taken the NSI parameters with their best-fit values from the combined analysis of NO$\nu$A and T2K. More specifically, $|\epsilon_{e\mu}|$ = 0.1, $\phi_{e\mu}$ = 1.06$\pi$ and $|\epsilon_{e\tau}|$= 0.033, $\phi_{e\tau}$ = 1.87$\pi$. With the inclusion of dual NSI, the allowed region corresponding to the higher octant disappears, and we are left only with the allowed region from the lower octant. 
\vspace{5mm}
In Fig. \ref{fig:image2} (middle panel), similarly, we display the allowed regions in the plane spanned by the standard CP-phase $\delta_{CP}$ and the atmospheric mixing angle $\theta_{23}$ in the NO case for T2HK. 
Comparing the SM scenario (left) with that of SM+2NSI (right) arising  simultaneously from $e-\mu$ and $e-\tau $sectors, the allowed region corresponding to the lower octant disappears, and we are left only with the allowed region from the higher octant. 
\vspace{5mm}
In Fig. \ref{fig:image2} (bottom panel), similarly, we display the allowed regions in the plane spanned by the standard CP-phase $\delta_{CP}$ and the atmospheric mixing angle $\theta_{23}$ in the NO case but now for a combination of DUNE and T2HK. Comparing the SM scenario (left) with that of the SM+2NSI scenario, the allowed region corresponding to the higher octant disappears and we are left only with the allowed region from the lower octant.  Comparing all three cases, i.e., DUNE, T2HK, and a combination of DUNE and T2HK, the allowed region corresponding to one of the octants disappears, and we are left only with the allowed region from the other octant.
\vspace{5mm}
Concerning the $\theta_{23}$ octant, we note that in the SM and SM+2NSI case arising from both the sectors, there is a clear preference for lower octant for DUNE ($\Delta\chi^{2}= 0.95$) and similarly for DUNE+T2HK ($\Delta\chi^{2}= 0.17$), whereas in case of T2HK there is a clear preference for higher octant ($\Delta\chi^{2}= 0.12$) where $\Delta\chi^2=\chi^{2}_{SM}-\chi^{2}_{SM+2NSI}$. Corresponding one-dimensional projection plots for $\delta_{CP}$ (left) and $\sin^{2}\theta_{23}$ (right) are displayed in Fig. \ref{fig:image3}. The contour plots depicting the allowed region corresponding to $\theta_{23}$ and $\delta_{CP}$ for DUNE and T2HK are displayed in Figs \ref{fig:image6} and \ref{fig:image7}.
\begin{figure}[htb]
\minipage{0.44\textwidth}
  \includegraphics[width=7.2cm,height=7.0cm]{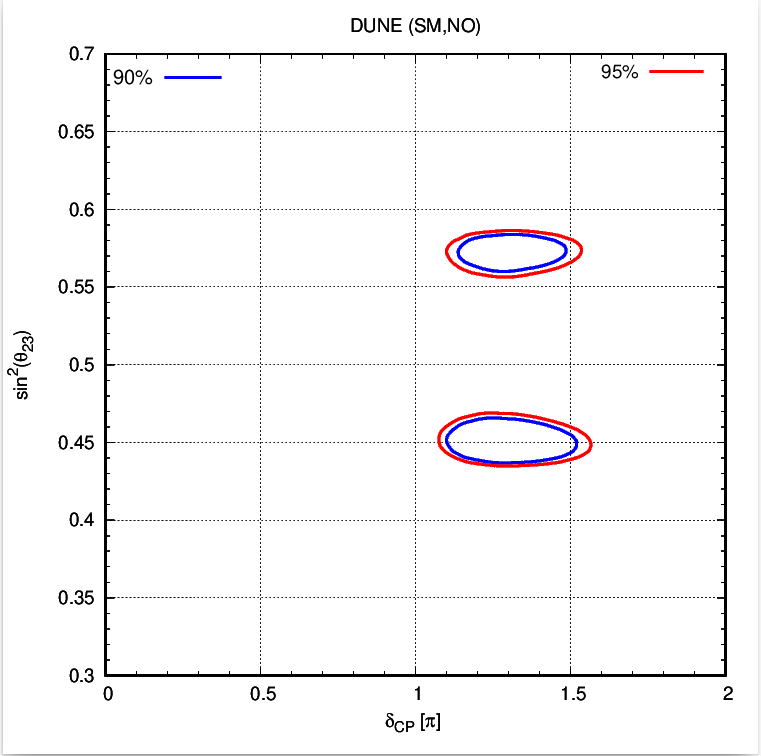}
\endminipage\hfill
\minipage{0.44\textwidth}
  \includegraphics[width=7.2cm,height=7.0cm]{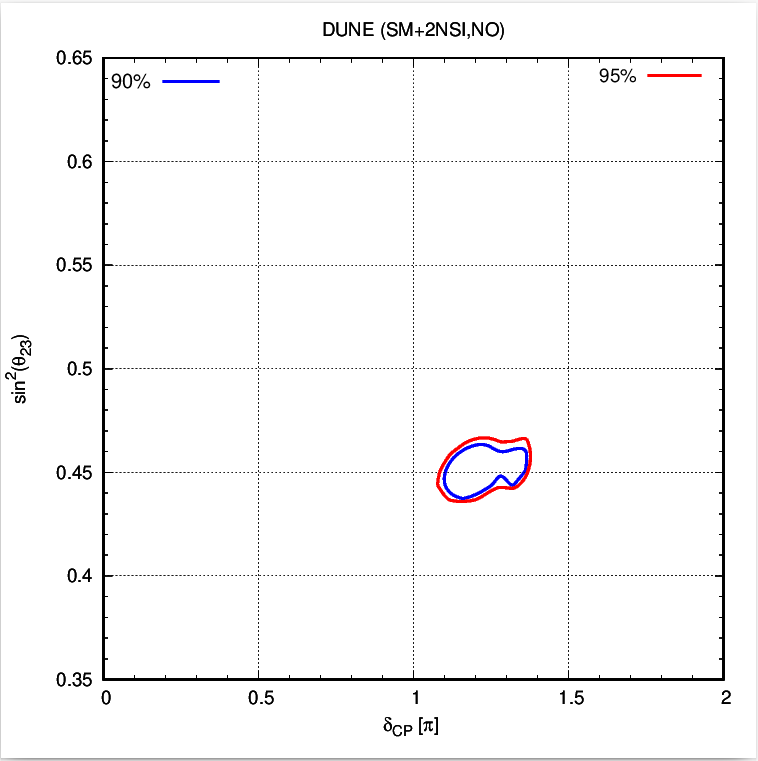}
\endminipage

\minipage{0.44\textwidth}
  \includegraphics[width=7.2cm,height=7.0cm]{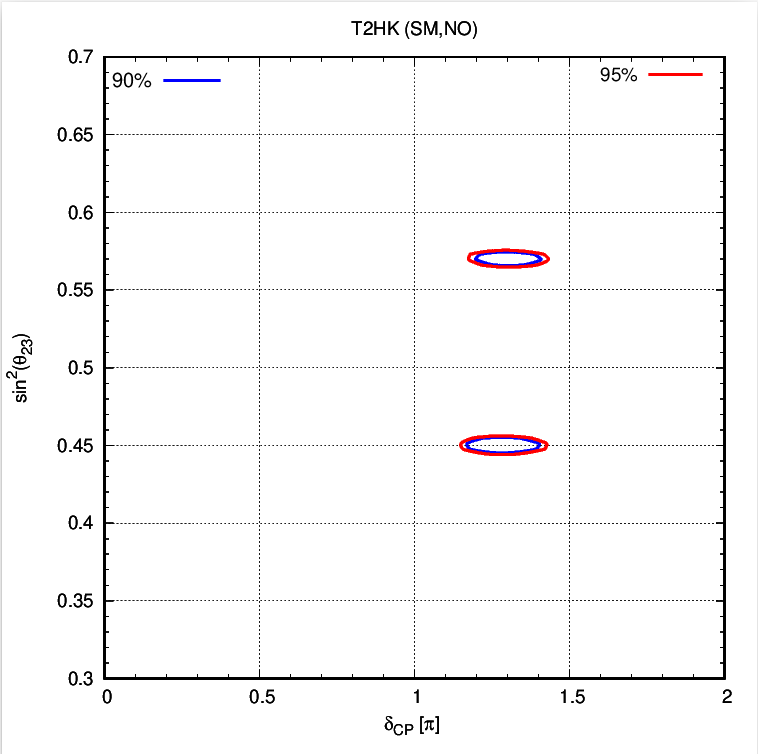}
\endminipage\hfill
\minipage{0.44\textwidth}
  \includegraphics[width=7.2cm,height=7.0cm]{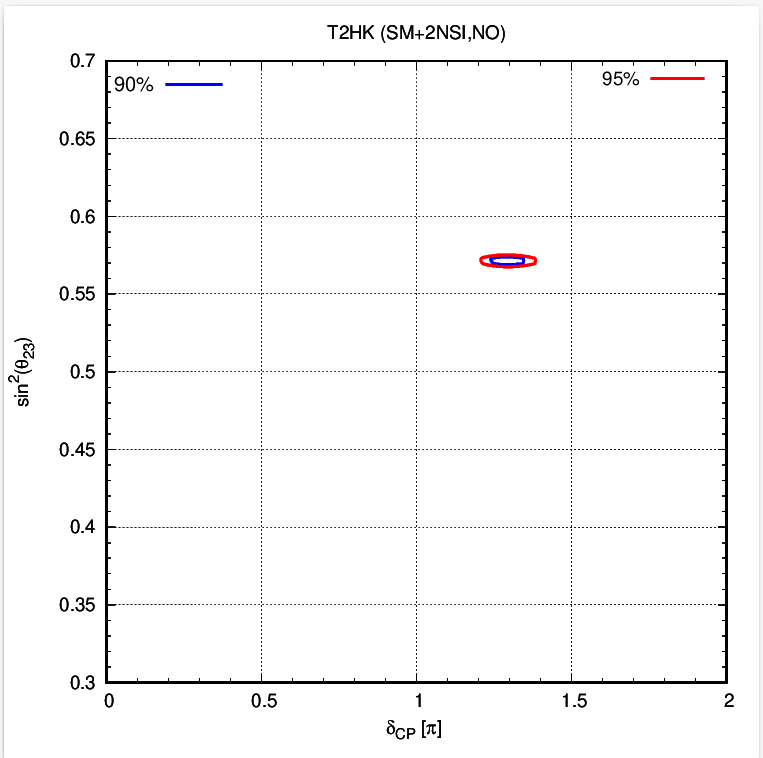}
\endminipage

\minipage{0.44\textwidth}
  \includegraphics[width=7.2cm,height=7.0cm]{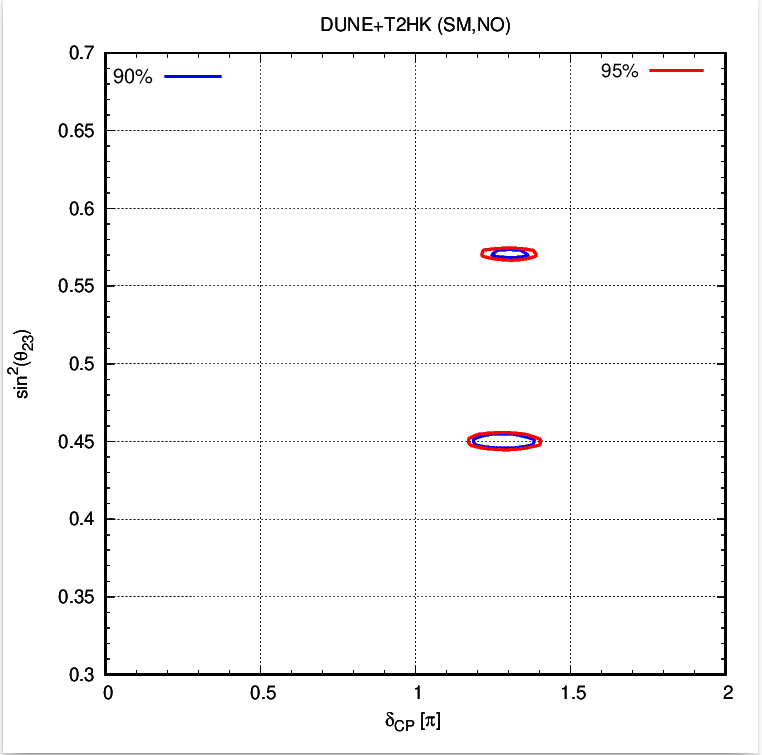}
\endminipage\hfill
\minipage{0.44\textwidth}
  \includegraphics[width=7.2cm,height=7.0cm]{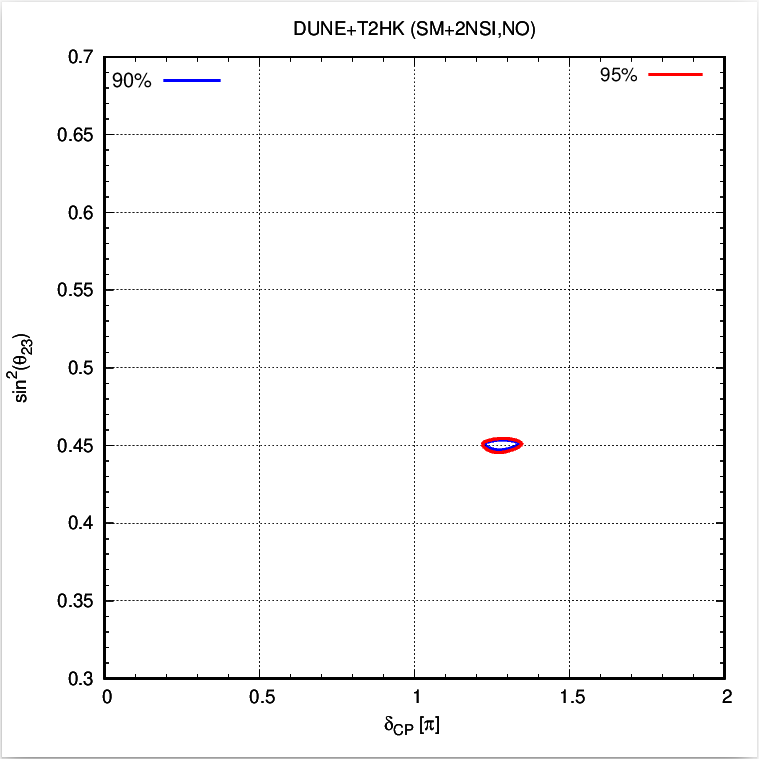}
\endminipage
\caption{Allowed regions determined separately by DUNE (top panel), T2HK (middle panel), and a combination of DUNE and T2HK (bottom panel)  for NO in the SM case (left panel) and with dual NSI arising from $e-\mu$ and $e-\tau$ sector (right panel). In the right panel, we have taken the NSI parameters at their best-fit values of NO$\nu$A+T2K ($|\epsilon_{e\mu}|$ = 0.1, $|\phi_{e\mu}|$ = 1.06$\pi$, $|\epsilon_{e\tau}|$ = 0.033, and $|\phi_{e\tau}|$ = 1.87$\pi$).
The contours are drawn at the 90$\%$ and 95$\%$ C.L. for 2 d.o.f.}\label{fig:image2}
\end{figure}

\begin{figure}[htb]
\minipage{0.44\textwidth}
  \includegraphics[width=7.2cm,height=7.0cm]{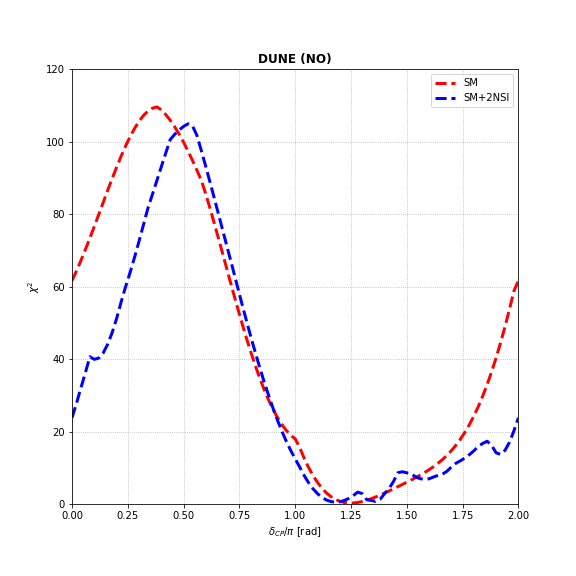}
\endminipage\hfill
\minipage{0.44\textwidth}
  \includegraphics[width=7.2cm,height=7.0cm]{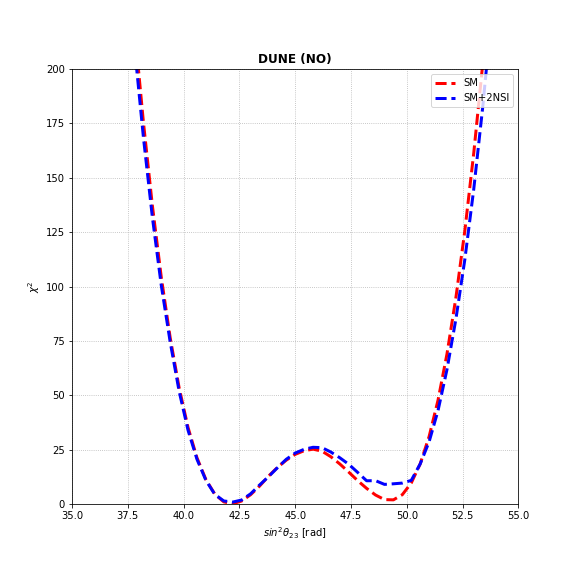}
\endminipage

\minipage{0.44\textwidth}
  \includegraphics[width=7.2cm,height=7.0cm]{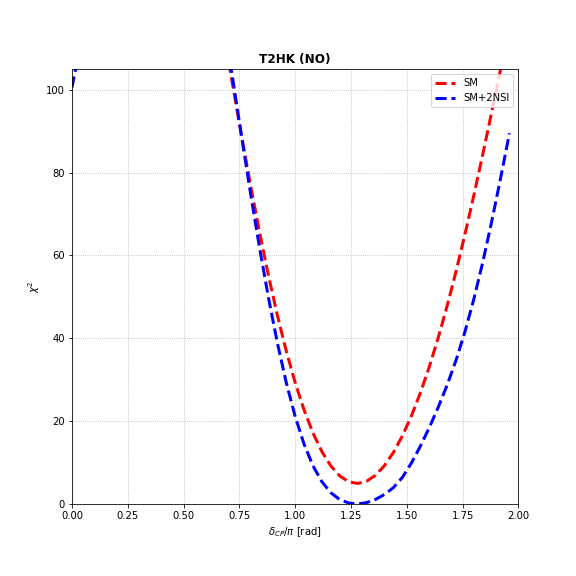}
\endminipage\hfill
\minipage{0.44\textwidth}
  \includegraphics[width=7.2cm,height=7.0cm]{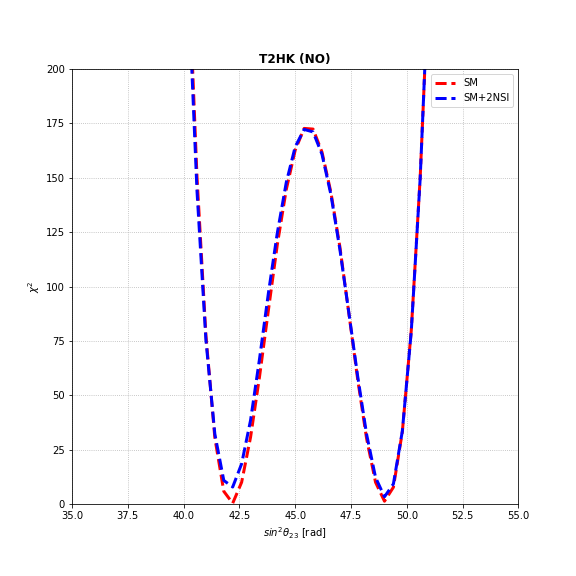}
\endminipage

\minipage{0.44\textwidth}
  \includegraphics[width=7.2cm,height=7.0cm]{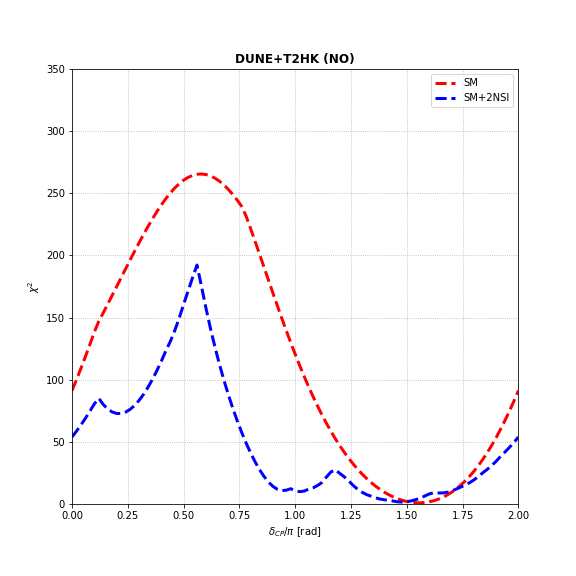}
\endminipage\hfill
\minipage{0.44\textwidth}
  \includegraphics[width=7.2cm,height=7.0cm]{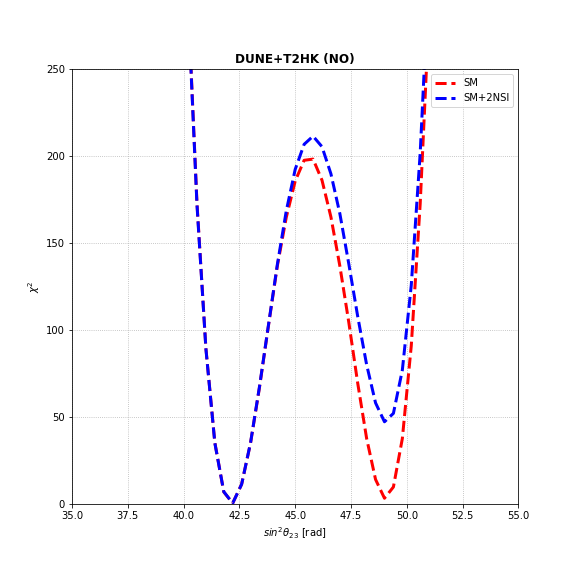}
\endminipage
\caption{One-dimensional projections of the standard parameters $\theta_{23}$ (left) and $\delta_{CP}$ (right) determined for DUNE (top panel), T2HK (middle panel) and DUNE+T2HK (bottom panel) in NO for SM
(red dashed curves) and SM+2NSI (blue dashed curves). }\label{fig:image3}
\end{figure}
\begin{figure}
\includegraphics[width=11.2cm,height=9.2cm]{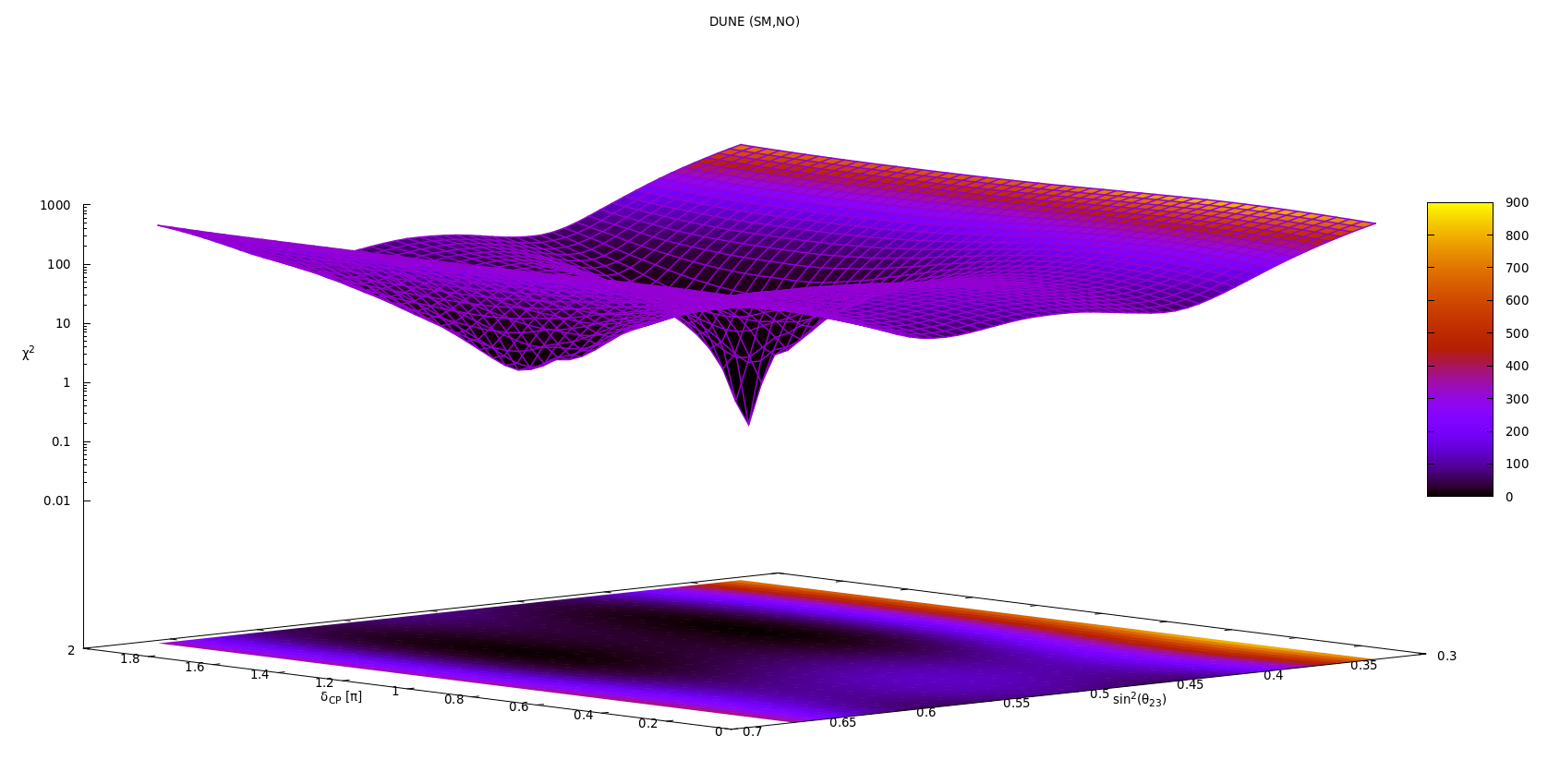}

\includegraphics[width=11.2cm,height=9.2cm]{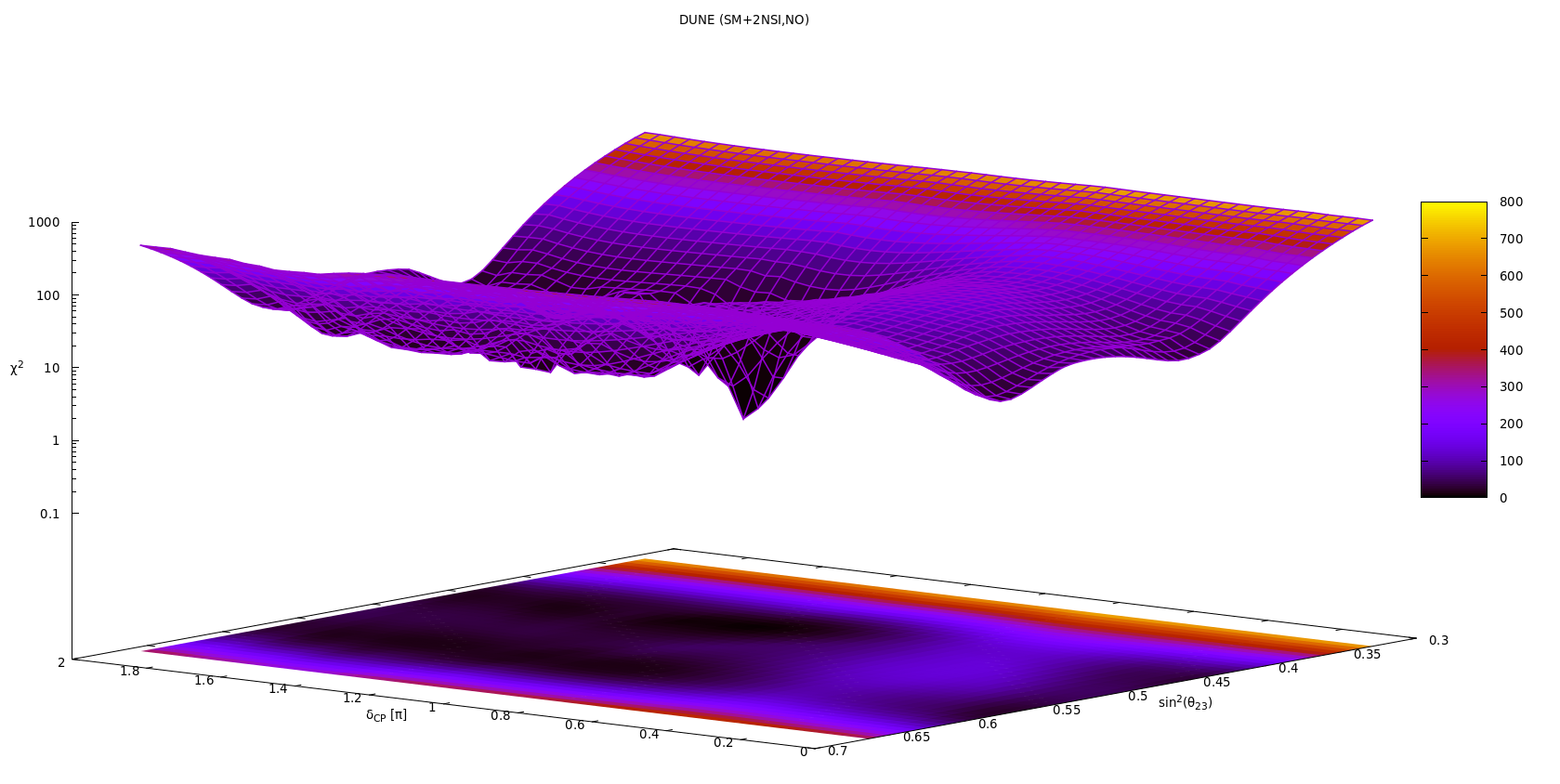}
\caption{Contour plots displaying allowed regions determined separately by DUNE for NO in the SM case (top) and with dual NSI arising from $e-\mu$ and $e-\tau$ sector (bottom)}\label{fig:image6}
\end{figure}
\begin{figure}
\includegraphics[width=11.2cm,height=9.2cm]{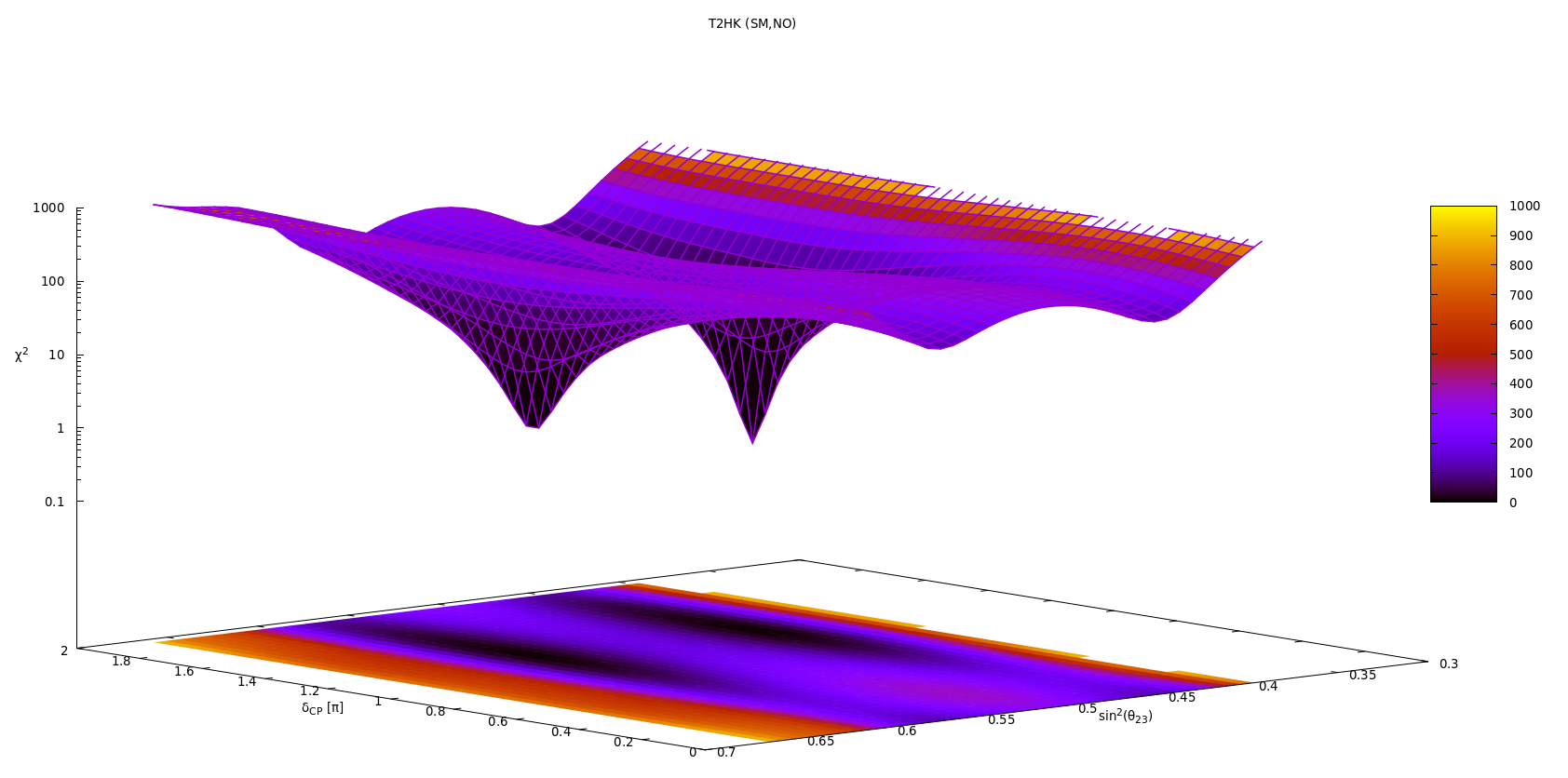}

\includegraphics[width=11.2cm,height=9.2cm]{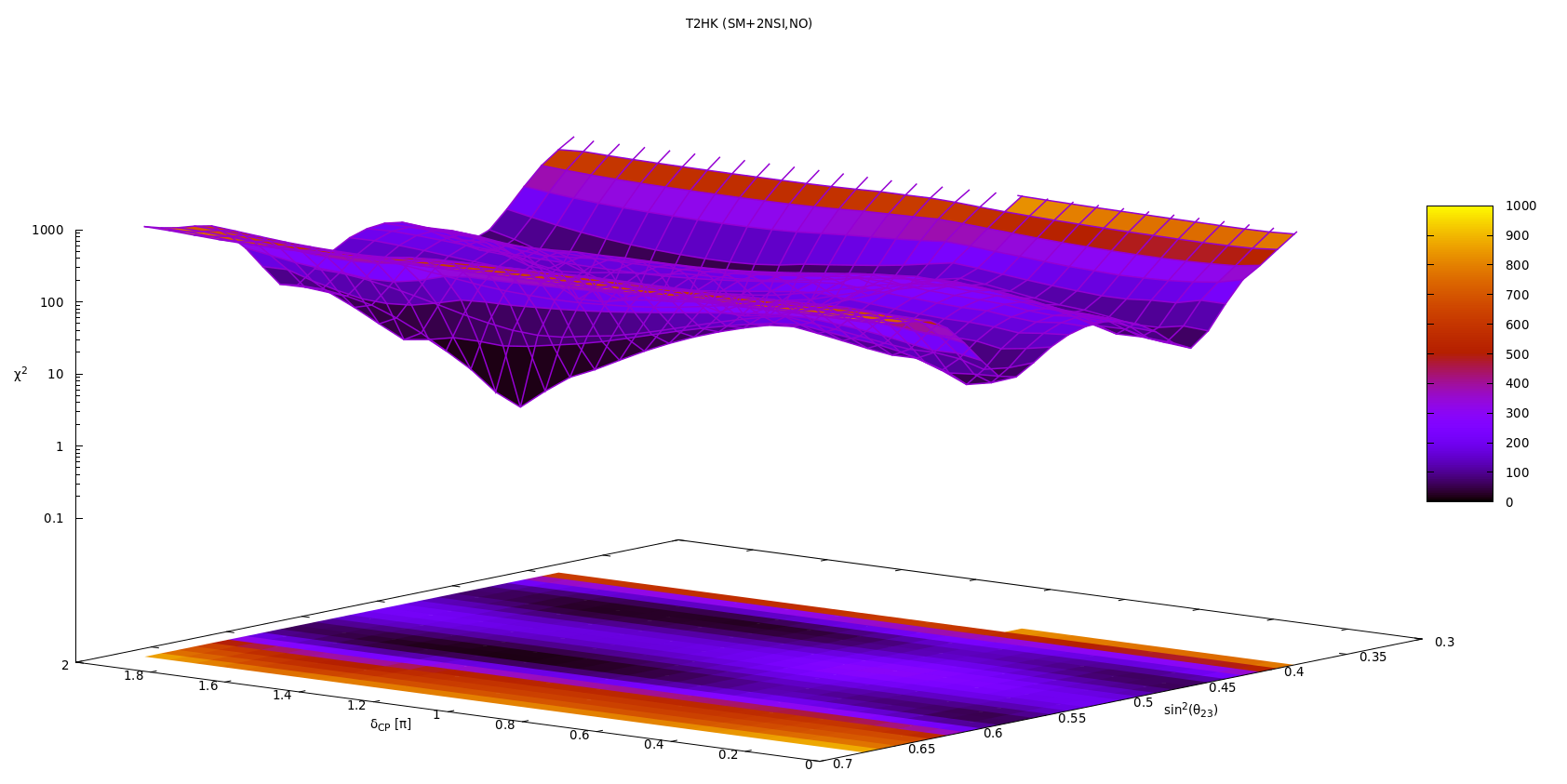}
\caption{Contour plots displaying allowed regions determined separately by T2HK for NO in the SM case (top) and with dual NSI arising from $e-\mu$ and $e-\tau$ sector (bottom)}\label{fig:image7}
\end{figure}}

\section{Effect of dual NSI Parameters on Oscillation Probability}
{For the upcoming long baseline experiments, T2HK and DUNE, it will be necessary to precisely measure the oscillation probabilities in order to determine the CP phase. However, when calculating these oscillation probabilities, the Earth's density uncertainty are taken into account in Ref \cite{King:2020ydu, Ghosh:2022bqj}.
Here, in this work, we have considered the effect of NSI on LBL experiments, DUNE, T2HK, and a combination of DUNE and T2HK, we discuss next the corresponding probability plots for both neutrino and anti-neutrino modes. In Fig. \ref{fig:image8} (top panel), the oscillation probability plots for DUNE in neutrino mode in the SM (left panel), SM+2NSI arising simultaneously from the $e-\mu$ and the $e-\tau$ sector (right panel) are shown. We see a good separation between NO-IO for both $\delta_{CP}=90^\circ$ and $\delta_{CP}=-90^\circ$ in the SM scenario. For the SM+2NSI scenario, we still have some separation between NO-IO for $\delta_{CP}=90^\circ$ in the mid-energy region. Whereas in the case of $\delta_{CP}=-90^\circ$ the NO-IO separation continuously decreases and they gradually merge around 4 GeV. For the anti-neutrino scenario, we see a reasonable separation between NO-IO for $\delta_{CP}=90^\circ$, 
 and $\delta_{CP}=-90^\circ$, for both SM and SM+2NSI case.
\begin{figure}[hbt!]
\minipage{0.45\textwidth}
\includegraphics[width=\linewidth]{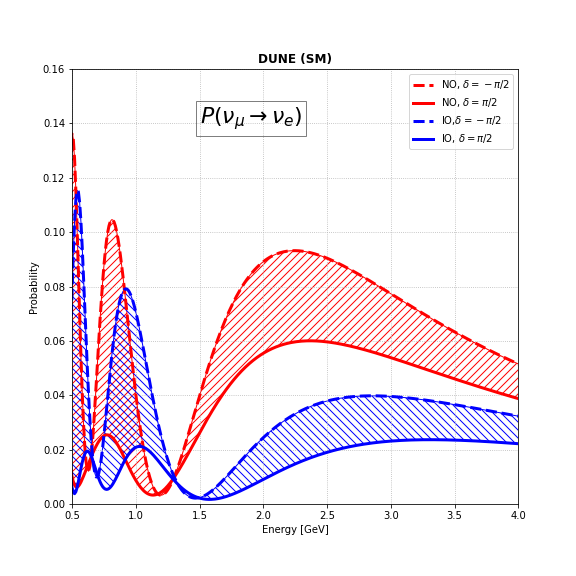}
\endminipage\hfill
\minipage{0.45\textwidth}
\includegraphics[width=\linewidth]{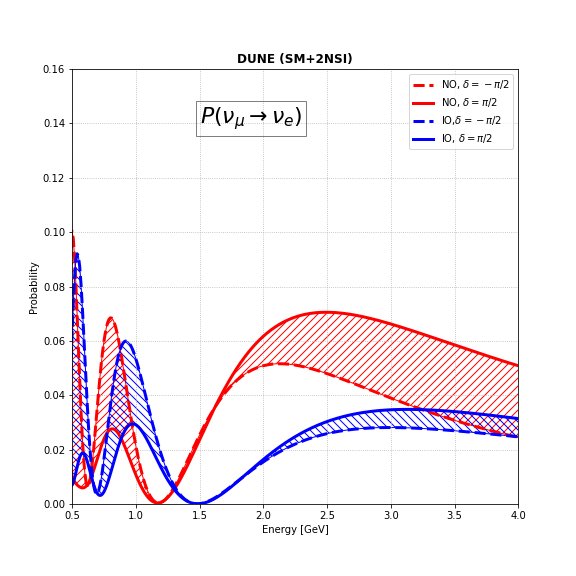}
\endminipage

\minipage{0.45\textwidth}
\includegraphics[width=\linewidth]{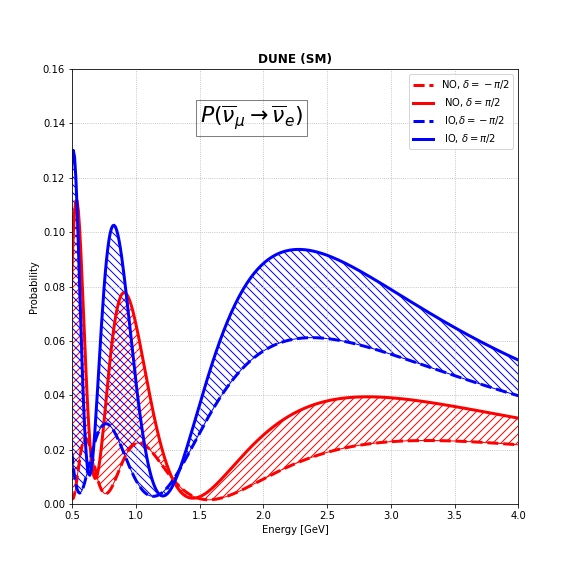}
\endminipage\hfill
\minipage{0.45\textwidth}
\includegraphics[width=\linewidth]{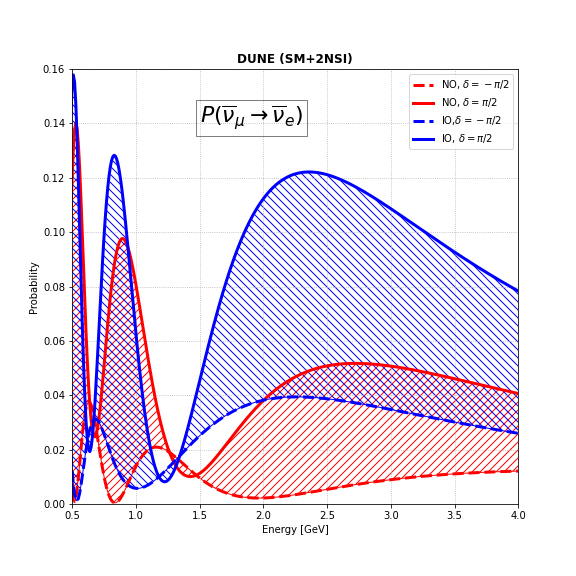}
\endminipage
\caption{Probability Plots for DUNE in SM (left) and SM+2NSI scenario with NSI arising from both $e-\mu$ sector and $e-\tau$ sector (right) for $\nu$ (top panel) and $\bar{\nu}$ (bottom panel) mode}\label{fig:image8}
\end{figure}

 In Fig. \ref{fig:image9} (top panel), the oscillation probability plots for T2HK in neutrino mode in the SM (left panel), SM+2NSI arising simultaneously from the $e-\mu$ and the $e-\tau$ sector (right panel) are shown. We visualize no separation between NO-IO for $\delta_{CP}=-90^\circ$ and a feeble separation between NO-IO for $\delta_{CP}=90^\circ$ in the SM scenario after 0.6 GeV energy. With the inclusion of dual NSI, we could perceive a feeble separation between NO-IO for $\delta_{CP}=-90^\circ$ till 0.7 GeV energy and better separation in case of $\delta_{CP}=90^\circ$ in the mid-energy region. We have repeated the exercise for the anti-neutrino case in T2HK which is displayed in Fig. \ref{fig:image9} (bottom panel).

\begin{figure}[hbt!]
\minipage{0.45\textwidth}
\includegraphics[width=\linewidth]{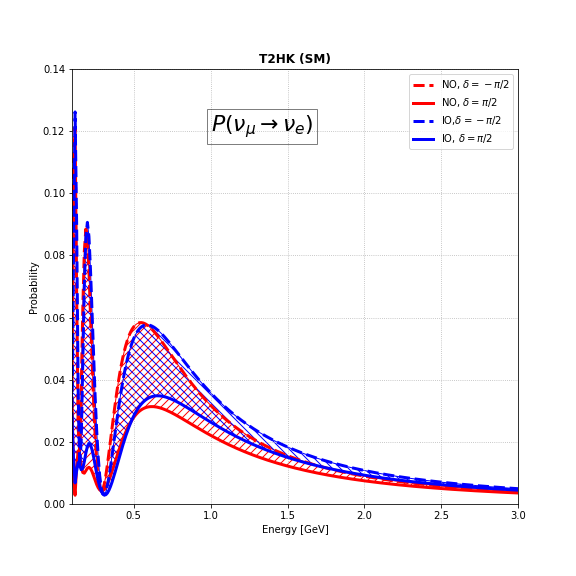}
\endminipage\hfill
\minipage{0.45\textwidth}
\includegraphics[width=\linewidth]{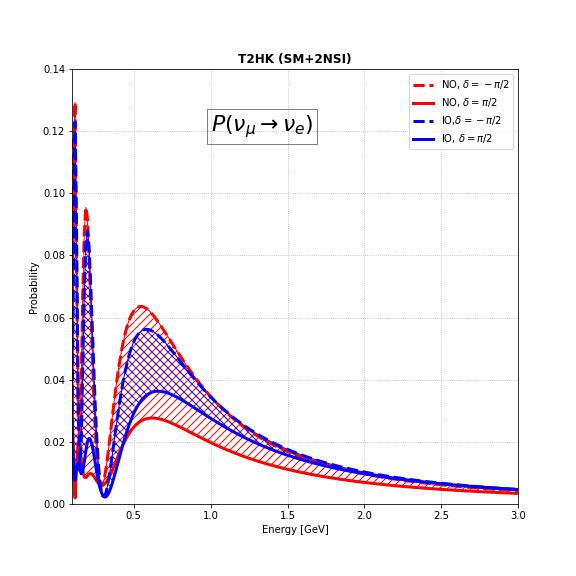}
\endminipage

\minipage{0.45\textwidth}
\includegraphics[width=\linewidth]{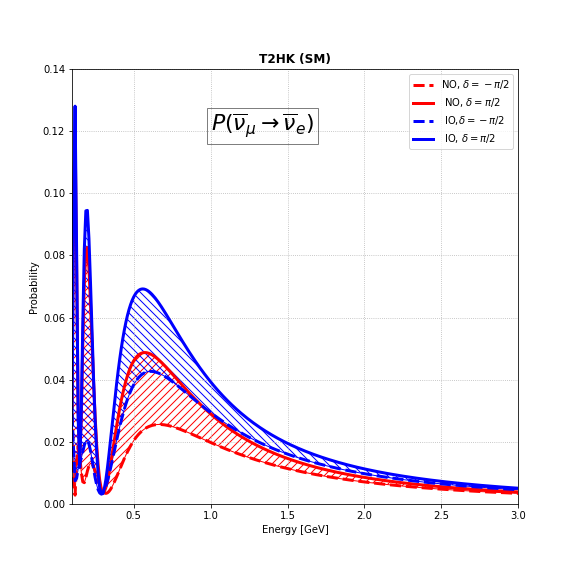}
\endminipage\hfill
\minipage{0.45\textwidth}
\includegraphics[width=\linewidth]{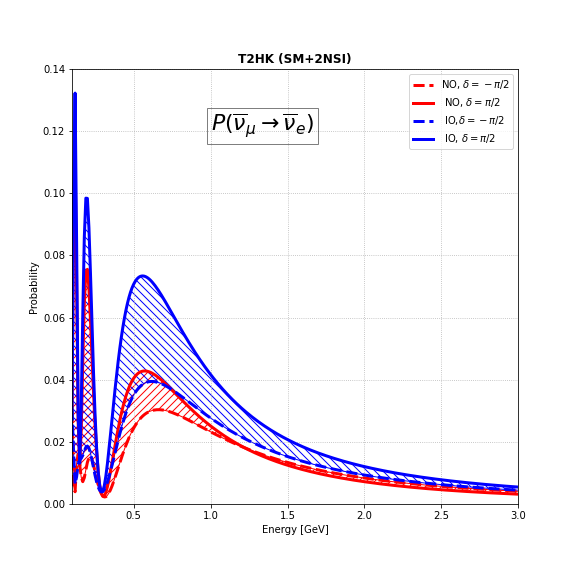}
\endminipage
\caption{Probability Plots for T2HK in SM (left) and SM+2NSI scenario with NSI arising from both $e-\mu$ sector and $e-\tau$ sector (right) for $\nu$ (top panel) and $\bar{\nu}$ (bottom panel) mode}\label{fig:image9}
\end{figure} 

In Fig. \ref{fig:image10} (top panel), the oscillation probability plots for a combination of DUNE and T2HK are displayed. The neutrino mode in the SM  case (left panel), and SM+2NSI arising simultaneously from the $e-\mu$ and the $e-\tau$ sector (right panel) are shown. We see a good separation between NO-IO for both $\delta_{CP}=90^\circ$ and $\delta_{CP}=-90^\circ$ in the SM scenario. For the SM+2NSI scenario, we have some separation between NO-IO for $\delta_{CP}=90^\circ$ in the mid-energy region till they merge around 2.7 GeV. Whereas, in the case of $\delta_{CP}=-90^\circ$ the NO-IO separation remains distinct. For the anti-neutrino scenario, we see a reasonable separation between NO-IO for $\delta_{CP}=90^\circ$. 
 and $\delta_{CP}=-90^\circ$, for SM case. With the inclusion of dual NSI, there is a continuous decrease in NO-IO separation for $\delta_{CP}=-90^\circ$ and they finally merge around 3.4 GeV. Whereas, for $\delta_{CP}=90^\circ$ the separation still remains distinct.
\begin{figure}[h!]
\minipage{0.45\textwidth}
\includegraphics[width=\linewidth]{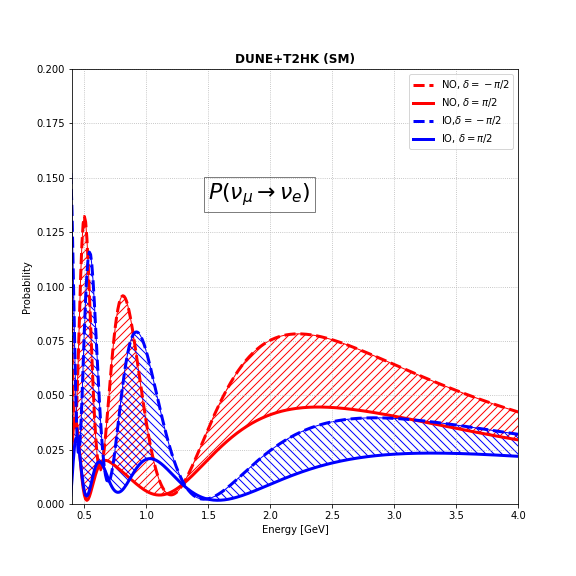}
\endminipage\hfill
\minipage{0.45\textwidth}
\includegraphics[width=\linewidth]{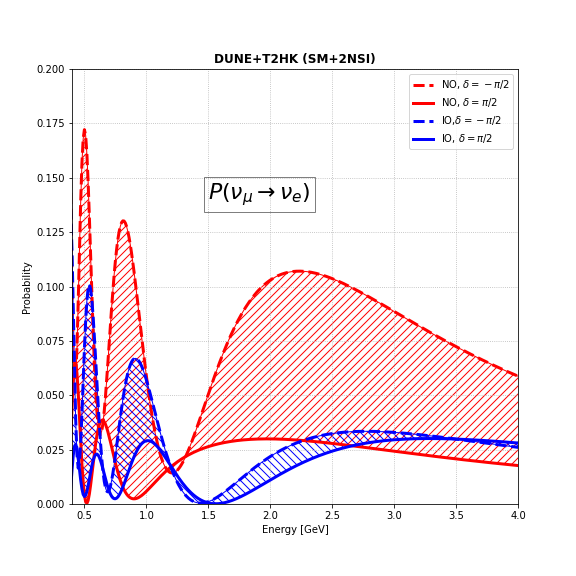}
\endminipage

\minipage{0.45\textwidth}
\includegraphics[width=\linewidth]{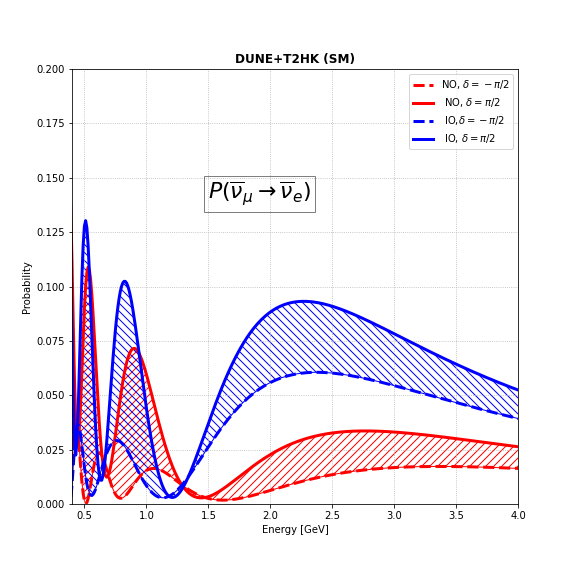}
\endminipage\hfill
\minipage{0.45\textwidth}
\includegraphics[width=\linewidth]{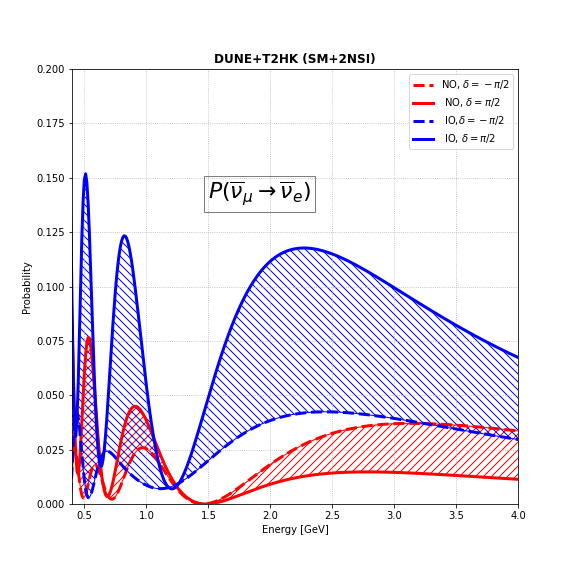}
\endminipage
\caption{Probability Plots for DUNE+T2HK in SM (left) and SM+2NSI scenario with NSI arising from both $e-\mu$ sector and $e-\tau$ sector (right) for $\nu$ (top panel) and $\bar{\nu}$ (bottom panel) mode}\label{fig:image10}
\end{figure}}

\section{CP Asymmetry}
{Here, we attempt to investigate the potential impact of dual NSIs on the CP measurement potential of two upcoming long-baseline studies, DUNE and T2HK. At the probability level, the CP-asymmetry parameter can be written as follows ~\cite{Medhi:2022qmu}:
\begin{equation*}
    A_{CP} = \frac{P(\nu_{\mu} \rightarrow \nu_{e})-P(\Bar{\nu_{\mu}} \rightarrow \Bar{\nu_{e}})}{P(\nu_{\mu} \rightarrow \nu_{e})+P(\Bar{\nu_{\mu}} \rightarrow \Bar{\nu_{e}})}
\end{equation*}

where, $P_{\mu e}$ and $\overline{P}{_{\mu e}}$ are the appearance probabilities of $\nu_{e}$ and $\overline{\nu}{_{e}}$ respectively. The CP asymmetry parameter, ($A_{CP}$), can be used to assess CP violation since it measures the change in oscillation probabilities when the CP phase changes sign. The shape and magnitude of the CP-asymmetry curve are strongly influenced by the baseline and energy. For DUNE and T2HK experiments the baselines are taken to be 1300km and 295km, respectively, whereas, energy are considered to be 2.6 GeV  and 0.6 GeV, respectively. In Fig. {\ref{fig:image11}}, one can visualize the significant separation between the CP asymmetry parameter for SM and SM with the inclusion of dual NSIs in T2HK than DUNE, for both NO and IO scenarios.

\begin{figure}[h!]
\minipage{0.45\textwidth}
\includegraphics[width=\linewidth]{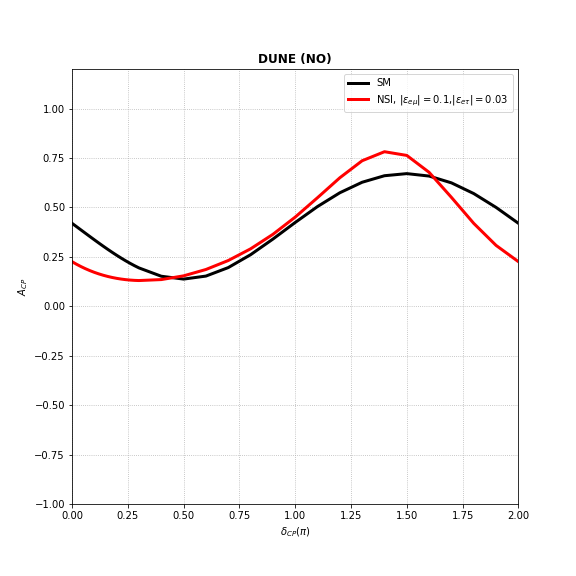}
\endminipage\hfill
\minipage{0.45\textwidth}
\includegraphics[width=\linewidth]{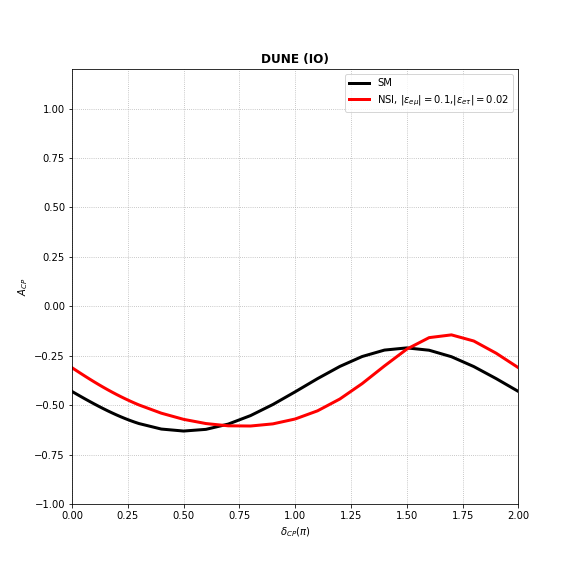}
\endminipage

\minipage{0.45\textwidth}
\includegraphics[width=\linewidth]{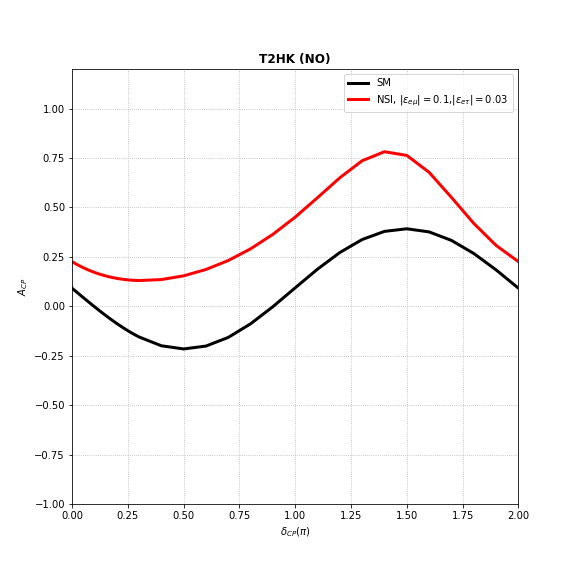}
\endminipage\hfill
\minipage{0.45\textwidth}
\includegraphics[width=\linewidth]{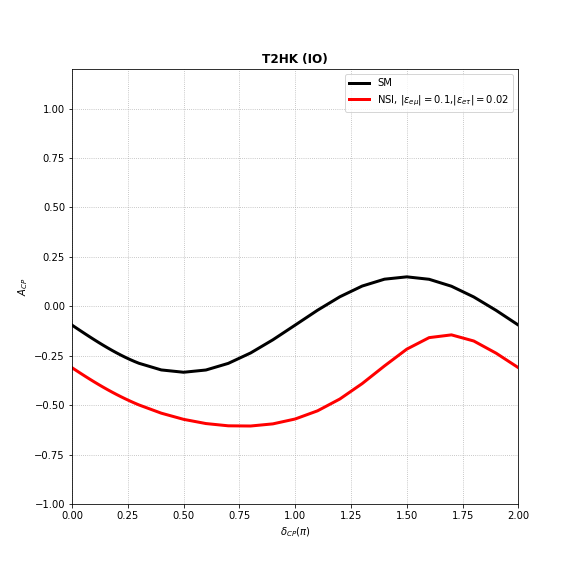}
\endminipage
\caption{CP asymmetry Plots for DUNE (top) and T2HK (bottom) in neutrino case for SM and SM+2NSI scenario with NSI arising from both $e-\mu$ sector and $e-\tau$ sector for NO (left) and IO (right) scenario.}\label{fig:image11}
\end{figure}}

\section{CP Violation Sensitivity}

{As mentioned before, one of the important objectives of the current and future LBL neutrino experiments is to determine the CP phase $\delta_{CP}$, as precisely as possible. In this context, the detection of CP-violation could play an important part in explaining the baryon asymmetry of the universe ~\cite{Cohen:1997ac,Fong:2012buy}. In the standard framework of three neutrino oscillations, we discuss here CP violation sensitivity. The signal indicating CP violation in the lepton sector ~\cite{Masud:2016bvp, Barger:2001yr, Burguet-Castell:2002ald} will be seen if the true values of $\delta_{CP}$ differ from the CP conserving values by a considerable amount ~\cite{Meloni:2018xnk}. Here,
\begin{figure}
\minipage{0.44\textwidth}
\includegraphics[width=7.2cm,height=7.0cm]{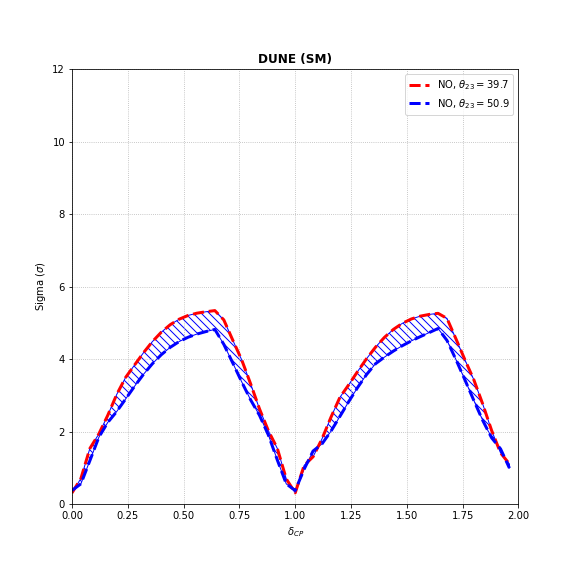}
\endminipage\hfill
\minipage{0.44\textwidth}
\includegraphics[width=7.2cm,height=7.0cm]{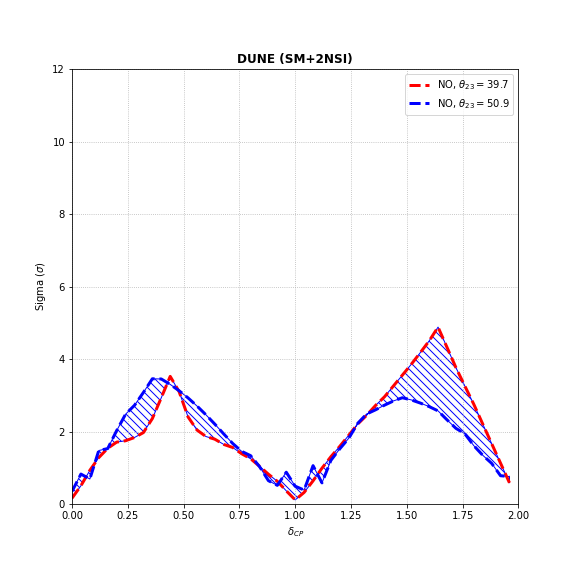}
\endminipage


\minipage{0.44\textwidth}
\includegraphics[width=7.2cm,height=7.0cm]{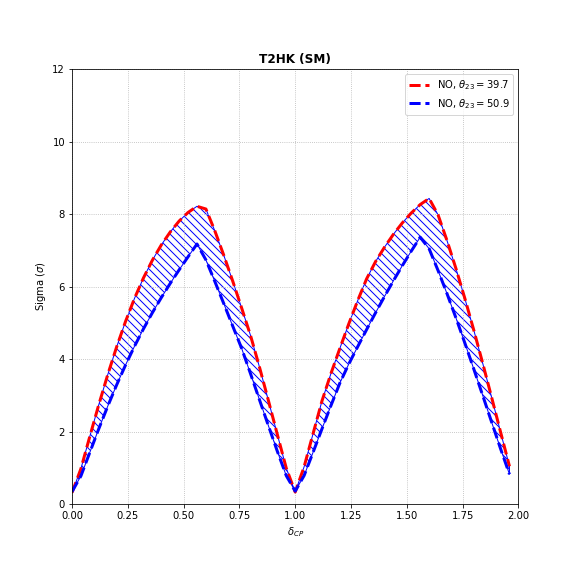}
\endminipage\hfill
\minipage{0.44\textwidth}
\includegraphics[width=7.2cm,height=7.0cm]{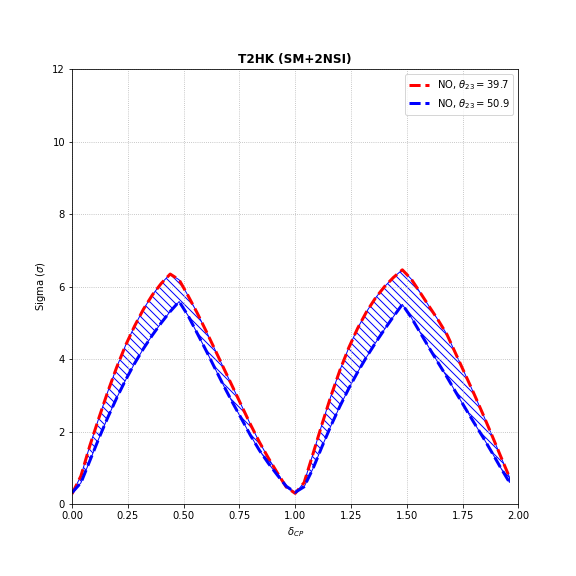}
\endminipage

\minipage{0.44\textwidth}
\includegraphics[width=7.2cm,height=7.0cm]{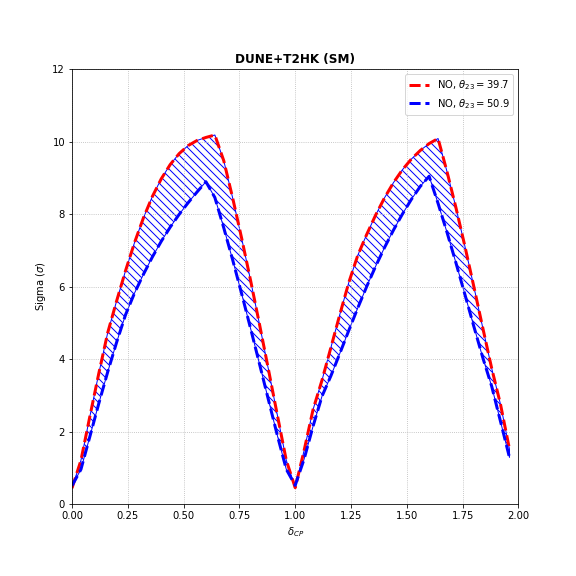}
\endminipage\hfill
\minipage{0.44\textwidth}
\includegraphics[width=7.2cm,height=7.0cm]{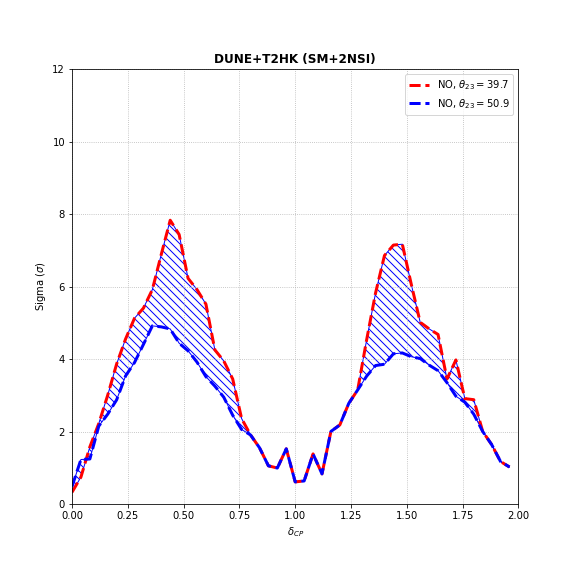}
\endminipage

\caption{CP discovery potential for DUNE (top panel), T2HK (middle panel), and DUNE+T2HK (bottom panel) as a function of the true value of the leptonic CP phase for NO in SM scenario(left panel) and SM+2NSI scenario (right panel). The bands represent the range in sensitivity obtained under the two different assumptions of $\theta_{23}$ value.}\label{fig:image12}
\end{figure}
$$\Delta \chi^{2}_{CPV} = Min[\Delta \chi^{2}_{CP}(\delta^{test}_{CP}=0), \Delta \chi^{2}_{CP}(\delta^{test}_{CP}=\pi)].$$

We find that for both DUNE and T2HK, as shown in Fig.\ref{fig:image12}, there are appreciable differences in the sensitivities for the SM+2NSI case in comparison to SM prediction. Moreover, in the case of the DUNE, there appears to be better sensitivity to NSI than in T2HK.}
\section{Plausible NSI Scenario}
{Here we consider a possible theoretical scenario in which similar kinds of NSI couplings, as obtained in our analysis, can be envisaged. The SM particle content is enlarged by two scalar fields, one SU(2)$_{L}$ doublet $\eta$ and one charged singlet $\phi$. In order to give rise to NSI, the SU(2)$_{L}$ doublet $\eta$ couples to the SM leptons via a renormalizable operator. In the presence of Z$_{2}$ under which $\eta$ and $\phi$ and the right-handed electron e$_{R}$ are odd, the only allowed term is \cite{Forero:2016ghr}:
\begin{equation}
     \mathcal{L} \supset \lambda_{\alpha}\overline{L_{\alpha}}\eta e_{R} +h.c.,
\end{equation}

where $\alpha$ is the flavor index, representing e, $\mu$ and $\tau$.

After integrating out heavy $\eta$, the effective operator can be rewritten as:

\begin{equation}
    \Delta \mathcal{L} = \frac{\lambda_{\alpha}\lambda^{*}_{\beta}}{m_{\eta}^{2}}(\overline{\nu_{\alpha}}\gamma^{\mu}\nu_{\beta})(\overline{e_{R}}\gamma_{\mu}e_{R})
\end{equation}

Recalling the standard expression for the neutral current non-standard interaction Lagrangian, which is generally represented by

\begin{equation}
    \mathcal{L} = -2\sqrt{2}G_{F}\epsilon_{\alpha\beta}^{fC}(\overline{\nu_{\alpha}}\gamma^{\mu}P_{L}\nu_{\beta})(\overline{f}\gamma_{\mu}P_{C}f)
\end{equation}

where $\alpha,\beta= e, \mu, \tau$ indicate the neutrino flavor, $f=u,e,d$ represent the matter fermions, the superscript $C=L,R$ refers to the chirality of the fermion current and $\epsilon_{\alpha\beta}$ are the effective NSI couplings. 

Comparing eqns (4) and (5), the relation between the NSI parameter $\epsilon_{\alpha\beta}$ and doublet $\eta$  can be written as

\begin{equation}
    \epsilon_{\alpha\beta}^{\eta} = - \frac{\lambda_{\alpha}\lambda^{*}_{\beta}}{4\sqrt{2}m_{\eta}^{2}G_{F}}
\end{equation}

After taking into account the $\eta-\phi$ mixing, the NSI from the two mass eigenstates s$_{1}$ and s${_2}$ are

\begin{equation}
    \epsilon_{\alpha\beta}^{s_{1}} = - \frac{\lambda_{\alpha}\lambda^{*}_{\beta}\theta^{2}}{4\sqrt{2}m_{1}^{2}G_{F}}
\end{equation}

\begin{equation}
    \epsilon_{\alpha\beta}^{s_{2}} = - \frac{\lambda_{\alpha}\lambda^{*}_{\beta}}{4\sqrt{2}m_{2}^{2}G_{F}}
\end{equation}

The magnitude of NSI can now be estimated from s$_{1}$ and s$_{2}$. 
\vspace{5mm}
\textbf{$e-\mu$ Sector}

For the $e-\mu$ sector, eqn (7) and eqn (8) can be rewritten as:
\begin{equation}
    \epsilon_{e\mu}^{s_{1}} = - \frac{\lambda_{e}\lambda^{*}_{\mu}\theta^{2}}{4\sqrt{2}m_{1}^{2}G_{F}}
\end{equation}
\begin{equation}
    \epsilon_{e\mu}^{s_{2}} = - \frac{\lambda_{e}\lambda^{*}_{\mu}}{4\sqrt{2}m_{2}^{2}G_{F}}
\end{equation}
In order to obtain sizable NSI, we need to measure the order of $\lambda_{e}$, $\lambda_{\mu}$ and $m_{2}$, which can be derived by imposing the bounds from $\mu \rightarrow 3e$ measurement. $\eta^{0}$ will induce $\mu^{-} \rightarrow e^{+}e^{-}e^{-}$ decay and the decay width normalised to the W-mediated  $\mu^{-} \rightarrow e^{-}\nu_{e}\overline{\nu_{\mu}} $.
\begin{equation}
    \frac{\Gamma_{\mu \rightarrow 3e}}{\Gamma_{\mu \rightarrow e\nu\overline{\nu}}} = \bigg|\frac{\lambda_{e}\lambda_{\mu}}{4\sqrt{2}m_{2}G_{F}}\bigg|^{2} \leq \frac{1.0 \times 10^{-12}}{1.0}
\end{equation}

\begin{equation}
    \frac{\sqrt{\lambda_{e}\lambda_{\mu}}}{m_{2}} \leq \frac{8.12 \times10^{-3}}{TeV}
\end{equation}

\textbf{$e-\tau$ Sector}:\\
Now for the $e-\tau$ sector, eqn (7) and eqn (8) can be rewritten as:

\begin{equation}
    \epsilon_{e\tau}^{s_{1}} = - \frac{\lambda_{e}\lambda^{*}_{\tau}\theta^{2}}{4\sqrt{2}m_{1}^{2}G_{F}}
\end{equation}

\begin{equation}
    \epsilon_{e\tau}^{s_{2}} = - \frac{\lambda_{e}\lambda^{*}_{\tau}}{4\sqrt{2}m_{2}^{2}G_{F}}
\end{equation}

In order to obtain sizable NSI, we need to measure the order of $\lambda$ and m$_{2}$, which can be derived by imposing the bounds from $\tau \rightarrow 3e$ measurement. $\eta^{0}$ will induce $\tau^{-} \rightarrow e^{+}e^{-}e^{-} $ decay and the decay width normalised to the W-mediated $\tau^{-} \rightarrow \mu^{-}\nu_{\tau}\overline{\nu_{\mu}} $, which is constrained by $\tau \rightarrow 3e$ results from Belle Collaboration ~\cite{Hayasaka:2010np}. 

\begin{equation}
    \frac{\Gamma_{\tau \rightarrow 3e}}{\Gamma_{\tau \rightarrow \mu\nu\overline{\nu}}} = \bigg|\frac{\lambda_{e}\lambda_{\tau}}{4\sqrt{2}m_{2}G_{F}}\bigg|^{2} \leq \frac{2.7 \times 10^{-8}}{0.17}
\end{equation}

which can be rewritten as:

\begin{equation}
    \frac{\sqrt{\lambda_{e}\lambda_{\tau}}}{m_{2}} \leq \frac{0.16}{TeV}
\end{equation}

Using eqn (12) and eqn (16), we obtain a relation between $\lambda_{\mu}$ and $\lambda_{\tau}$ as below:

\begin{equation}
   \sqrt{ \frac{\lambda_{e}\lambda_{\mu}}{\lambda_{e}\lambda_{\tau}}} = \frac{8.12 \times 10
   ^{-3} \hspace{3mm}( m_{2} \hspace{1mm}TeV)}{0.16 \hspace{3mm}(m_{2} \hspace{1mm}TeV)}
\end{equation}

\begin{equation}
    \lambda_{\mu} = 2.57 \times 10^{-3} \lambda_{\tau}
\end{equation}

\subsection{\Large{Constraint on $e-\mu$ Sector}}

In order to obtain the NSI constraint from the $e-\mu$ sector, the expression from eqn (9) and eqn (10) can be rewritten as:

\begin{equation}
    |\epsilon_{e\mu}^{s_{1}}| =  3.9 \times 10^{-4} \bigg(\frac{\theta}{m_{1}/m_{2}} \bigg)^{2} \bigg(\frac{\sqrt(\lambda_{e}\lambda_{\mu})}{8.12\times10^{-3}} \bigg)^{2}  \bigg(\frac{TeV}{m_{2}} \bigg)^{2}
\end{equation}

\begin{equation}
    |\epsilon_{e\mu}^{s_{2}}| =  3.9 \times 10^{-4}\bigg(\frac{\sqrt(\lambda_{e}\lambda_{\mu})}{8.12\times10^{-3}}\bigg)^{2} \bigg(\frac{TeV}{m_{1}} \bigg)^{2}
\end{equation}

The term $\frac{TeV}{m_{1}}$ in eqns (19) and (20)  are of order 1. Thus, the new form of equations are:

\begin{equation}
    |\epsilon_{e\mu}^{s_{1}}| =  3.9 \times 10^{-4} \bigg(\frac{\theta}{m_{1}/m_{2}} \bigg)^{2} \bigg(\frac{\sqrt(\lambda_{e}\lambda_{\mu})}{8.12\times10^{-3}} \bigg)^{2} 
\end{equation}

\begin{equation}
    |\epsilon_{e\mu}^{s_{2}}| =  3.9 \times 10^{-4}\bigg(\frac{\sqrt(\lambda_{e}\lambda_{\mu})}{8.12\times10^{-3}}\bigg)^{2}
\end{equation}

Assuming $\lambda_{e} \sim \lambda_{\mu} = 1$ and the ratio of the mixing angle between $\eta-\phi$ coupling $\theta$  and masses $m_{1}$ and $m_{2}$, i.e., $\frac{\theta}{m_{1}/m_{2}}$ to be of the order of 0.1.

\begin{equation}
    |\epsilon_{e\mu}^{s_{1}}| \leq  0.59 
\end{equation}

\begin{equation}
    |\epsilon_{e\mu}^{s_{2}}| \leq  5.9
\end{equation}

Considering both the contributions from s$_{1}$ and s$_{2}$, the NSI parameter is found to be $\epsilon_{e\mu} \leq 3.245$

\subsection{\Large{Constraint on $e-\tau$ Sector}}

In order to obtain the NSI constraint from the $e-\tau$ sector, the expression from eqn (13) and eqn (14) can be rewritten as:
\begin{equation}
    |\epsilon_{e\tau}^{s_{1}}| =  3.9 \times 10^{-4} \bigg(\frac{\theta}{m_{1}/m_{2}} \bigg)^{2} \bigg(\frac{\sqrt(\lambda_{e}\lambda_{\tau})}{8.12\times10^{-3}} \bigg)^{2}  \bigg(\frac{TeV}{m_{2}} \bigg)^{2}
\end{equation}

\begin{equation}
    |\epsilon_{e\tau}^{s_{2}}| =  3.9 \times 10^{-4}\bigg(\frac{\sqrt(\lambda_{e}\lambda_{\tau})}{8.12\times10^{-3}}\bigg)^{2} \bigg(\frac{TeV}{m_{1}} \bigg)^{2}
\end{equation}

The term $\frac{TeV}{m_{1}}$ in eqns (25) and (26)  are of order 1. Thus, the new form of equations are:

\begin{equation}
    |\epsilon_{e\tau}^{s_{1}}| =  3.9 \times 10^{-4} \bigg(\frac{\theta}{m_{1}/m_{2}} \bigg)^{2} \bigg(\frac{\sqrt(\lambda_{e}\lambda_{\tau})}{8.12\times10^{-3}} \bigg)^{2} 
\end{equation}

\begin{equation}
    |\epsilon_{e\tau}^{s_{2}}| =  3.9 \times 10^{-4}\bigg(\frac{\sqrt(\lambda_{e}\lambda_{\tau})}{8.12\times10^{-3}}\bigg)^{2}
\end{equation}

Similar to the $e-\mu$ sector, $\lambda_{e} \sim \lambda_{\mu} = 1$  and from eqn (18) $\lambda_{\tau} = 389$.\\
The ratio of the mixing angle between $\eta-\phi$ coupling $\theta$  and masses $m_{1}$ and $m_{2}$, i.e., $\frac{\theta}{m_{1}/m_{2}}$ to be of the order of 0.1.

\begin{equation}
    |\epsilon_{e\tau}^{s_{1}}| \leq  0.59 
\end{equation}

\begin{equation}
    |\epsilon_{e\tau}^{s_{2}}| \leq  5.9
\end{equation}

Considering both the contributions from s$_{1}$ and s$_{2}$, the NSI parameter can be obtained as $\epsilon_{e\tau} \leq 3.245$.\\
The NSI constraints obtained using our GLoBES analysis for the experimental setup and the combined datasets of NO$\nu$A, and T2K are of similar orders as predicted by the theoretical scenario, discussed above, of the SM extended by two scalar fields, one doublet, and one charged singlet. It should also be noted here that the extracted NSI parameters in our exercise are compatible with existing constraints ~\cite{ IceCubeCollaboration:2021euf}.

\section{Conclusions}
{In this article, we assumed that new physics occurs in the form of NSI contributing simultaneously from $e-\mu$ and $e-\tau$ sectors. In doing so, we obtained the constraints on NSI parameters by combining the NO$\nu$A and T2K datasets. Thereafter, we used the derived constraints and have shown 
that for $\theta_{23}$, when we use NSI arising from both the sectors, simultaneously, DUNE prefers the lower octant, T2HK prefers the higher octant, and a combination of both DUNE and T2HK prefers the lower octant. One-dimensional projection plots and three-dimensional contour plots depict the same outcome. Moreover, we observed the striking effects of dual NSI constraints on both neutrino and anti-neutrino channel probabilities in DUNE, T2HK, and a combination of DUNE and T2HK. In turn, this can help us to understand the neutrino mass ordering problem. Here, in our work, the assessment of CP asymmetry reveals that with the inclusion of NSI from $e-\mu$ and $e-\tau$ sectors, simultaneously, we visualize a significant separation between SM and SM+2NSI in T2HK than in DUNE. Furthermore, the CP discovery potential showed that the effect of dual NSIs reduces the sensitivity, which is prominent in DUNE in comparison to T2HK, and a combination of both. Further studies will help us to
understand the nature of NSI and help extract the parameters in the neutrino sector, including unambiguous determination of the leptonic CP phase $\delta_{CP}$.}       
\nocite{*}
\section {Acknowledgement}
\large{The authors acknowledge the support from the DST, India under project No. SR/MF/PS-01/2016-IITH/G and fellowship from  MoE, India.}

\bibliographystyle{my-JHEP}
\bibliography{references}  

 
\end{document}